# Revealing the Partially Coherent Nature of Transport in IGZO

Ying Zhao[1*], Michiel J. van Setten[1], Anastasiia Kruv[1], Pietro Rinaudo[1,3], Harold Dekkers[1], Jacopo Franco[1], Ben Kaczer[1], Mischa Thesberg[2], Gerhard Rzepa[2], Franz Schanovsky[2], Attilio Belmonte[1], Gouri Sankar Kar[1], Adrian Chasin[1]

[1]imec, Leuven, Belgium
[2]Global TCAD Solutions (GTS) GmbH, Vienna, Austria
[3]KULeuven, Leuven, Belgium
*Email: ying.zhao@imec.com

## Abstract

Thin-film transistors based on amorphous oxide semiconductors (AOS) are promising candidates for enabling further DRAM scaling and 3D integration, which are critical for advanced computing. Despite extensive research, the charge transport mechanism in these disordered semiconductors remains poorly understood. In this work, we investigate charge transport in the archetypical AOS material, indium gallium zinc oxide (IGZO), across a range of compositions and temperatures using thin-film transistors and Hall bar structures. Our results show that the electrons involved in transport exhibit partially spatial coherence and non-degenerate conduction. Under these conditions, transport is dominated by electron transfer across insulating gaps between locally coherent regions, rather than by degenerate percolative transport above a mobility edge, or by localized-state hopping, both of which are widely assumed in the literature. To describe this behavior, we develop a field-effect-aware fluctuation-induced tunnelling (FEAFIT) model. The model accurately captures experimental data across all compositions, temperatures, and gate voltages, and enables extraction of fundamental transport parameters. These tunnelling parameters are directly related to electron coherence dimensions and the degree of energetic disorder obtained from first-principles calculations, allowing them to serve as a physically interpretable descriptor of macroscopic conduction. Our findings advance the fundamental understanding of charge transport in AOS-based transistors and provide a foundation for further performance improvements.

## Introduction

Since the seminal work of Nomura[1], indium gallium zinc oxide (IGZO)-based thin-film transistors (TFTs) have found applications in displays[2,3] , flexible electronics[4], ubiquitous RFID tags[5], quasi-non-volatile DRAM[6–8], and other technologies essential for the CMOS 2.0 era[9]. These diverse applications stem from the unique properties of amorphous oxide semiconductors (AOS): low-temperature processing (enabling low-cost flexible and printed electronics as well as BEOL compatibility), excellent deposition uniformity (ensuring electrical consistency for large-area displays), low off-currents (providing long retention times in DRAM cells), and electron mobility comparable to crystalline phases. High on-current AOS-based transistors are expected to play a crucial role in achieving ultra-high-density storage devices[10,11]. A comprehensive understanding of charge transport in AOS is therefore essential for improving intrinsic mobility and advancing this field.

Remarkably, despite their disordered structure, AOS materials exhibit high electron mobility[1,6,8]. Unlike amorphous silicon (a-Si)[12,13] and various organic semiconductors[14,15], carrier coherence in AOS is not fully destroyed by atomic disorder, allowing carriers to behave as quasi-free waves within confined regions. For example, amorphous IGZO (a-IGZO) can exhibit Hall-effect mobilities at high carrier concentrations[10,16–18]. A Hall study on doped IGZO reported temperature-independent Hall carrier concentration ($n_{\mathrm{Hall}}$) alongside thermally activated DC conductivity, suggesting degenerate conduction[16]. However, classical Lorentz force considerations for Hall measurements apply only to fully delocalized carriers with well-defined drift velocities, not to localized carriers[19–21]. In materials with mixed localized and delocalized states, Hall measurements can be misleading, which is overlooked in the previous study[16]. Furthermore, some researchers have observed temperature-independent conductivity in IGZO at cryogenic temperatures[18,22], which contradicts traditional transport models such as Multiple Trapping and Release (MTR) or Variable Range Hopping (VRH)[23].

To reconcile the measured Hall mobility with the thermally activated conductivity in doped IGZO, Kamiya et al. proposed a spatially varying potential landscape with random barrier heights above a distinct mobility edge[16,24]. However, this percolation-based model is unsatisfactory because it assumes fully coherent electron transport and neglects the impact of energetic disorder, which leads to loss of coherence and the formation of localized states[25]. Consequently, it fails to reproduce mobility at low temperatures and cannot accurately describe the Seebeck coefficient[18,25]. More recently, Nenashev et al.[26] and Fishchuk et al.[27,28] introduced a percolation description based on a random band-edge model that accounts for spatial variations in the mobility edge position[29]. Yet, this approach still does not capture the observed low-temperature features. Lee et al.[30] attempted to combine multiple transport models to fit experimental data across wide gate-voltage and temperature ranges, but this method requires 13 fitting parameters and lacks fundamental physical insight. Beyond percolation models, Wang et al.[18] and Nenashev et al.[31] proposed field-induced lateral tunnelling to explain temperature-independent conductivity at cryogenic temperatures. While these models can reproduce certain experimental trends, they offer limited ability to extract deeper physical insights or establish direct links between film composition, microstructure, and transport properties.

To uncover the nature of transport in a-IGZO, we modulate the degree of energetic disorder and examine the coherence of electronic states by varying the material composition. Increasing the indium concentration enhances electron doping efficiency and effective mobility[1,10,32,33], whereas higher gallium content introduces greater energetic disorder along the transport paths[17,25,34]. Our results show that chemical composition significantly influences the temperature dependence of transport. Specifically, we observe a unique temperature-independent field-effect (FE) conductivity at cryogenic temperatures, which disappears as the Ga content increases. Hall measurements combined with first-principles calculations reveal that electrons remain partially coherent within finite regions, the size of which is determined by composition.

The presence of partial electron coherence implies quasi-free transport within these coherent regions, coupled with fluctuation-induced tunnelling between them. Building on this observation, we propose a new transport model that provides a unified and quantitative description of the temperature-dependent conductivity in amorphous IGZO-based transistors. This model remains valid across the entire experimental range of temperature and gate voltage. Furthermore, we extract the incoherence dimensionality for different a-IGZO compositions, and these physical parameters exhibit consistent trends with first-principles predictions. This agreement reinforces the partially coherent transport picture and validates our proposed model for a-IGZO.

# Results

### Temperature dependent transport measurements

We use back-gated a-IGZO TFTs, fabricated on 300 mm wafers, for transport studies (Fig. 1a). The optimized a-IGZO deposition by PVD guarantees film thickness uniformity across the wafer (deposited a-IGZO thickness map in Fig. S1b). As previously reported[32], the initial threshold voltage ($V_{th0}$) increases and the field-effect mobility ($\mu_{FE}$) decreases with increasing Ga content, at a fixed In content (~39%), as shown in Fig. 1b. The $\mu_{FE}$ also decreases with increasing Ga content, for fixed In-to-Ga ratio as shown in Fig. 1c. In addition, we observe that Ga-rich a-IGZO films have a larger band gap, as shown by the ellipsometry measurements in Fig. 1d. Assuming that the valence band remains constant due to the $O_{2p}$ metal d hybrids[1], we conclude that Ga-rich a-IGZO has a higher $E_{CBM}$ (referred to the vacuum level), revealing that the Ga concentration affects on the conduction states.

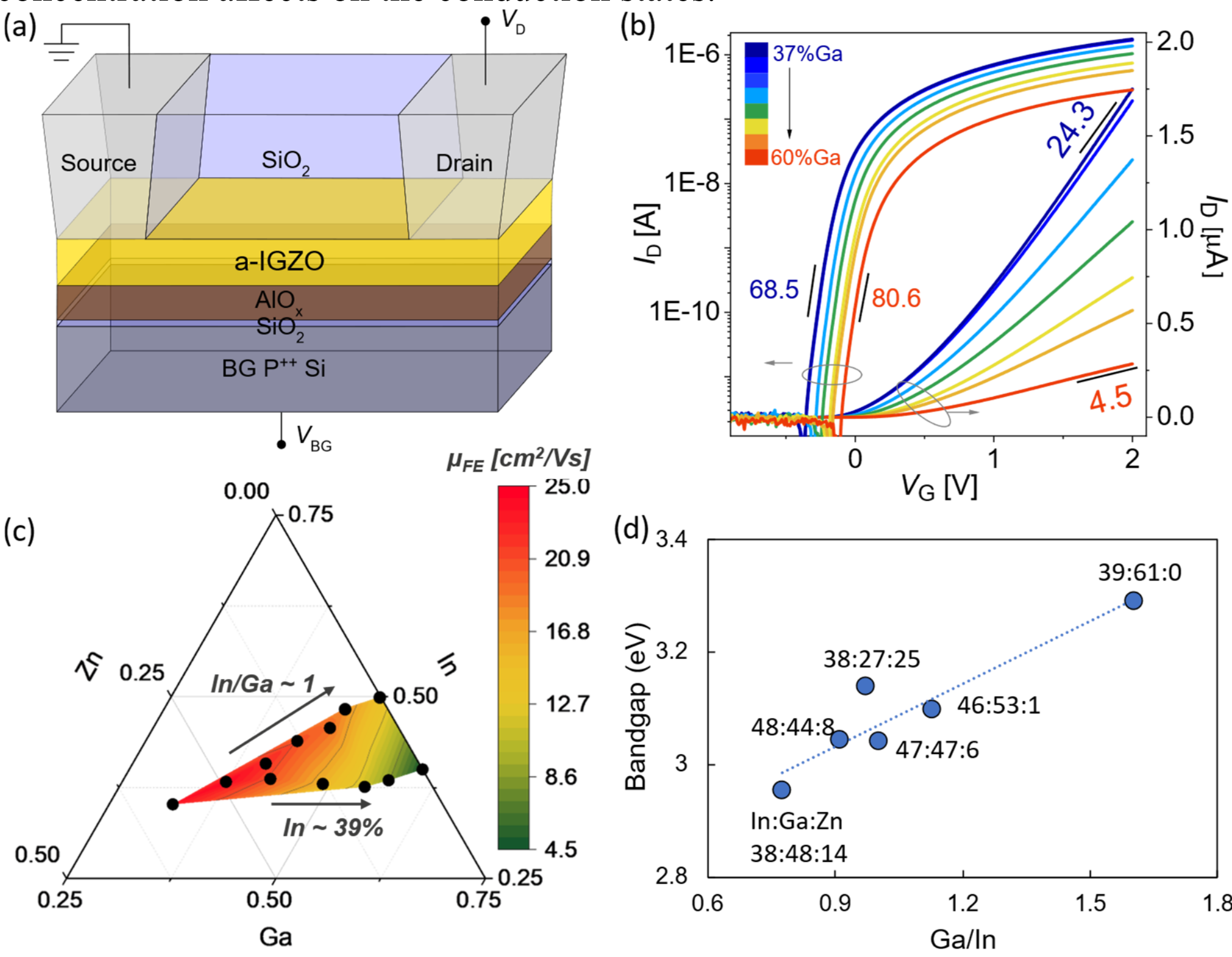

**Fig. 1 The fabricated a-IGZO TFTs and composition-related properties.** (a) Schematic of back-gated a-IGZO transistor's structure for transport measurement. (b) Transfer characteristics of IGZO TFTs with different compositions under room temperature, at $V_D$ = 0.05 V. (c) Ternary composition diagram with colour-mapped field effect mobility ($\mu_{FE}$) under two routes of varying IGZO compositions, fixed In concentration at 39 % and fixed In/Ga ratio at 1. (d) Band gap characterization by ellipsometry for different compositions, the ratio among In, Ga and Zn is marked.

We observe a strong impact of composition on the temperature-dependent conductivity ($\sigma$) characteristics. Fig. 2a shows the temperature dependence of conductivity at a fixed gate voltage ($V_G$ - $V_{FB}$) of three a-IGZO compositions, Ga37 (In:Ga:Zn = 39:37:24), Ga49 (In:Ga:Zn = 39:49:12) and Ga61 (In:Ga:Zn = 39:61:0). The flat band voltages ($V_{FB}$) are obtained from TCAD calibration in Note S2. Conductivity is thermally activated for all compositions and the temperature dependence increases with Ga content. We phenomenologically divide the experimental conductivity into three regions, where the conductivity follows a linear relation with temperature on a log scale. For the first two regimes, all compositions share the same temperature range: the first regime spans 300 K to 120 K, and the second spans 120 K to 50 K. Below 50 K, Ga37 gradually reaches a saturation state with temperature independence, while Ga61 continues to exhibit significant thermal activation down to 10 K. Ga49 exhibits moderate slopes of thermal activation, which saturate at high gate voltage (see Fig. S7). For each composition and regime, we fit an Arrhenius dependence to obtain the activation energy $E_a$ as a function of gate voltage. The extracted $E_a$'s in the different regimes are shown in Fig. 2b. The $E_a$ of Ga61 is more than twice than the $E_a$ of Ga37 in all regimes. Specifically, in the last regime for $T < 40$ K, $E_{a,3}$ in Ga61 is $10 - 20$ meV, whereas in Ga37, $E_{a,3}$ approaches zero. The gradual suppression of the temperature dependence cannot be accounted for by thermally activated hopping or trapping descriptions. At the same time, the finite activation energies ($E_{a,3} > 0$ eV) in Ga49 and Ga61 indicate that, electrons still experience energy disorder along their conducting paths, even when thermal activation is strongly suppressed.

To quantitatively characterize the energetic disorder associated with composition, we extract the tail density of states (DOS) from the gate- and temperature-dependent FE conductivity. As shown in Fig. 2c, increasing Ga concentration leads to a systematic broadening of the tail states, accompanied by a reduction of the density near the edge of fully delocalized states ($E_{dloc}$). A single exponential form cannot adequately capture the extracted distributions over the energy range. Instead, two characteristic slopes are needed, indicating the presence of multiple disorder scales. Detailed fitting procedures and parameters are discussed in the supplementary notes (Note S1). Importantly, this tail-state analysis implies transport not band-like. Even the high-density, narrow tail states near $E_{dloc}$ remain globally localized and incoherent.

Besides the activation energy and tail DOS, the exponent $\gamma$ of a power-law dependence between conductivity and the applied $V_G - V_{FB}$ (Fig. S8) provides an additional phenomenological indicator of disorder according to Vissenberg–Matters model[35,36]. The value of $\gamma$ decreases with temperature but increases with Ga concentration, as depicted in Fig. 2d. However, unlike the linear $\gamma \propto 1/T$ relation expected for hopping mechanism[36,37], the temperature dependence of $\gamma$ in cannot be described by a simple linear form. Moreover, the temperature dependence of $\gamma$ in a-IGZO deviates strongly from this behavior and exhibits the same three-regime structure observed in $\sigma(T)$. This further suggests that transport in a-IGZO cannot be consistently described by thermally activated hopping mechanisms of electrons near the Fermi level ($E_F$) alone.

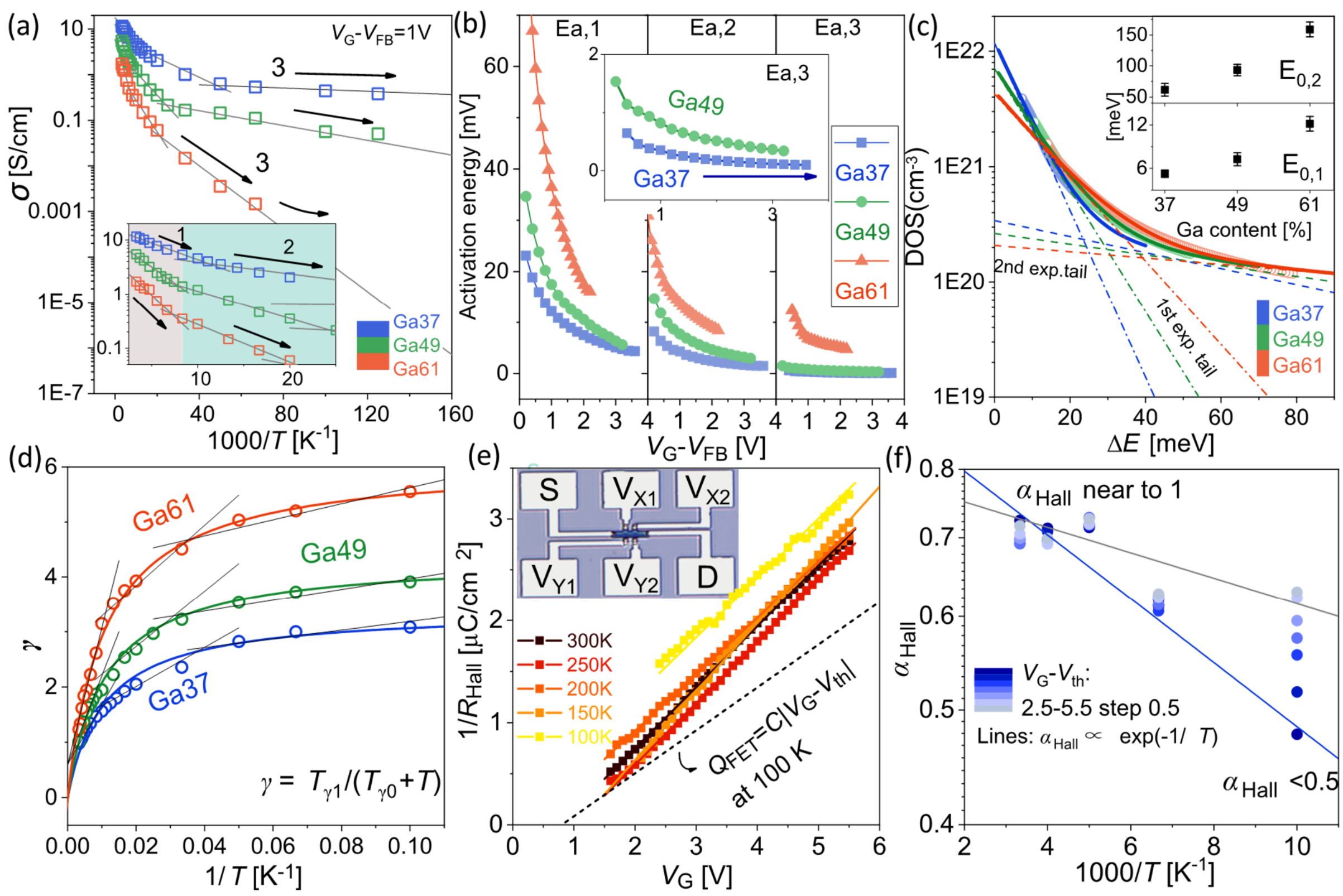

**Fig. 2 Unconventional transport behavior in IGZO with varying composition.** (a) Temperature dependent conductivity $\sigma$ in Arrhenius coordinates for three IGZO compositions: Ga37 (In:Ga:Zn = 39:37:24), Ga49 (In:Ga:Zn = 39:49:12) and Ga61 (In:Ga:Zn=39:61:0). $\sigma$ is divided into three regimes for Arrhenius fitting ($\sigma \propto \exp\left(-E_{a,(1,2,3)}/k_B T\right)$, gray lines), labeled as regime 1-3. The inset provides a magnified view from 50 K to 300 K. (b) The extracted $E_a$ of different Arrhenius fitting regimes. (c) The extracted density of tail states of three compositions. The solid lines represent fittings of two exponential tails $N_{\text{tail}}(E) = N_{0,1}\exp\left(-E/E_{0,1}\right) + N_{0,2}\exp\left(-E/E_{0,2}\right)$. Dash-dot lines correspond to the first exponential tail ($E_{0,1}$, $N_{0,1}$), while dashed lines represent the second tail ($E_{0,2}$, $N_{0,2}$). The insets show the composition dependence of the two tail widths. (d) Power-law exponent $\gamma$ extracted from the power-law fits, plotted against inverse temperature (1/$T$) from 10 K to 300 K. The black solid lines indicate linear fits ($\gamma = A_\gamma/T + B_\gamma$) based on VM model[35], while the solid-coloured lines correspond to the FEAFIT model fitting ($\gamma = T_{\gamma 1}/(T_{\gamma 0} + T)$). (e) Temperature dependence of inverse Hall coefficient 1/ $R_{\text{Hall}}$ as a function of $V_G$. The black dashed lines represent the charge density $Q_{\text{FET}} = C_{\text{ox}} \mid V_G - V_{\text{th}} \mid$. The inset shows an optical image of Hall-bar structure used for FE Hall measurement, with a channel width of 6 μm and length of 34 μm. (f) Temperature dependence of carrier coherence factor $\alpha_{\text{Hall}}$ under varying $V_G$ - $V_{\text{th}}$. The solid lines serve as a thermal activation guide for eye.

## Field effect Hall measurements

We now reveal the incoherent nature of transport and corroborate the extracted high density of tail states in a-IGZO by a Hall-effect measurement in field-effect transistors (FETs), which is a more rigorous method[19,20,40] than a direct Hall measurement in doped IGZO film[15,16]. AC Hall measurements were performed on the standard composition (Ga37) a-IGZO FETs using the Hall-bar structure shown in the inset of Fig. 2e. The temperature range used is from 300 K to 100 K, and additional measurement details, data analysis and extraction are provided in methods and supplementary notes. The temperature dependence of the inverse Hall coefficient (1/$R_{\text{Hall}}$) as a function of $V_G$ is shown in Fig. 2e. We observe the value of 1/$R_{\text{Hall}}$ is greater than the total electrostatically induced charge density $Q_{\text{FET}}$, which means the Hall carrier density ($n_{\text{Hall}}$) overestimates the total carrier density ($n_{\text{FET}}$). The overestimation of carrier density is due to the compensation of Hall voltage ($V_{\text{Hall}}$) by incoherent carriers that are pushed by the transverse field $V_{Hall}$, causing a drift in the opposite direction of the Lorentz force[20]. The difference becomes greater at lower temperatures, indicating an increase in contribution of incoherent carriers.

The partial coherence is parameterized by the carrier coherence factor $\alpha_{Hall}$, defined as $n_{FET}/n_{Hall}$[19]. Less coherent carriers lead to an underdeveloped Hall effect with $\alpha_{Hall} < 1$[19]. The value of $\alpha_{Hall}$ in our a-IGZO transistors is less than unity from 300 K to 100 K, as shown in Fig. 2f, indicating a globally incoherent transport[20]. At high temperature and gate bias, $\alpha_{Hall}$ is closer to 1 because $E_F$ is close to $E_{dloc}$ and carriers have a greater possibility of achieving degenerate conduction. At low biases and temperatures, $\alpha_{Hall}$ is less than 0.5 since fewer coherent carriers are present. The results of our Hall measurements indicate an absence of fully delocalized carriers and metallic conduction in the a-IGZO transistors. The "variable band edge" used in literature to introduce a metal-insulator transition energy range also does not provide a correct picture. The transport occurs within partially coherent states, characterized by narrow tail widths and high densities as we extracted in Fig. 2c, moving without a well-defined wave vector.

**First principles computations**

To further the understanding of the electron states of the different compositions, we perform first-principles computations within the density functional theory (DFT) framework. Per composition, five strongly elongated models of about 500 atoms were generated and carefully optimized to investigate the electronic states. The size of these models, approximately 13 nm in the long direction, allows us to resolve the localization of the conduction orbitals in at least one direction[41]. As shown in Fig. 3a, conduction localization is evident in all three structural models. However, we observe orbitals with an extension in the range of many atomic sites. The local density of states, shown in Figs. 3b–d, also reveals the locally coherent nature of isolated low-energy regions. The absence of a continuous connection between these regions, for reasonably accessible Fermi levels, prevents global coherence. We observe a broader spatial spread of orbitals around 8 nm for Ga37, two separated orbitals for Ga49, and a single, highly localized state of about 3 nm for Ga61. This trend indicates that increasing Ga content leads to stronger localization of the conduction states, reflecting reduced electron coherence and suppressed spatial delocalization. In contrast, most prior simulations of amorphous IGZO used cubic cells smaller than 2 nm per side, which are insufficient to capture conduction band localization, often leading to the incorrect conclusion of complete delocalization.

In field-effect transistors, $E_F$ is shifted by the gate voltage. As $E_F$ increases, more of the low-energy pockets are filled and the spread of electron wavepackets becomes larger. The energy height and spacing of the barriers between the coherent regions become lower and smaller, respectively. As the first-principles simulations use a quasi-1D geometry, the extracted barrier heights and widths may not match the absolute values seen in experiments on full 3D systems. In 3D, electrons can find lower barrier paths that are not present in a 1D cut. However, the relative trends across different compositions, such as increasing barrier separation with Ga content, and the underlying partially coherence-limiting picture are expected to hold in both 1D and 3D regimes.

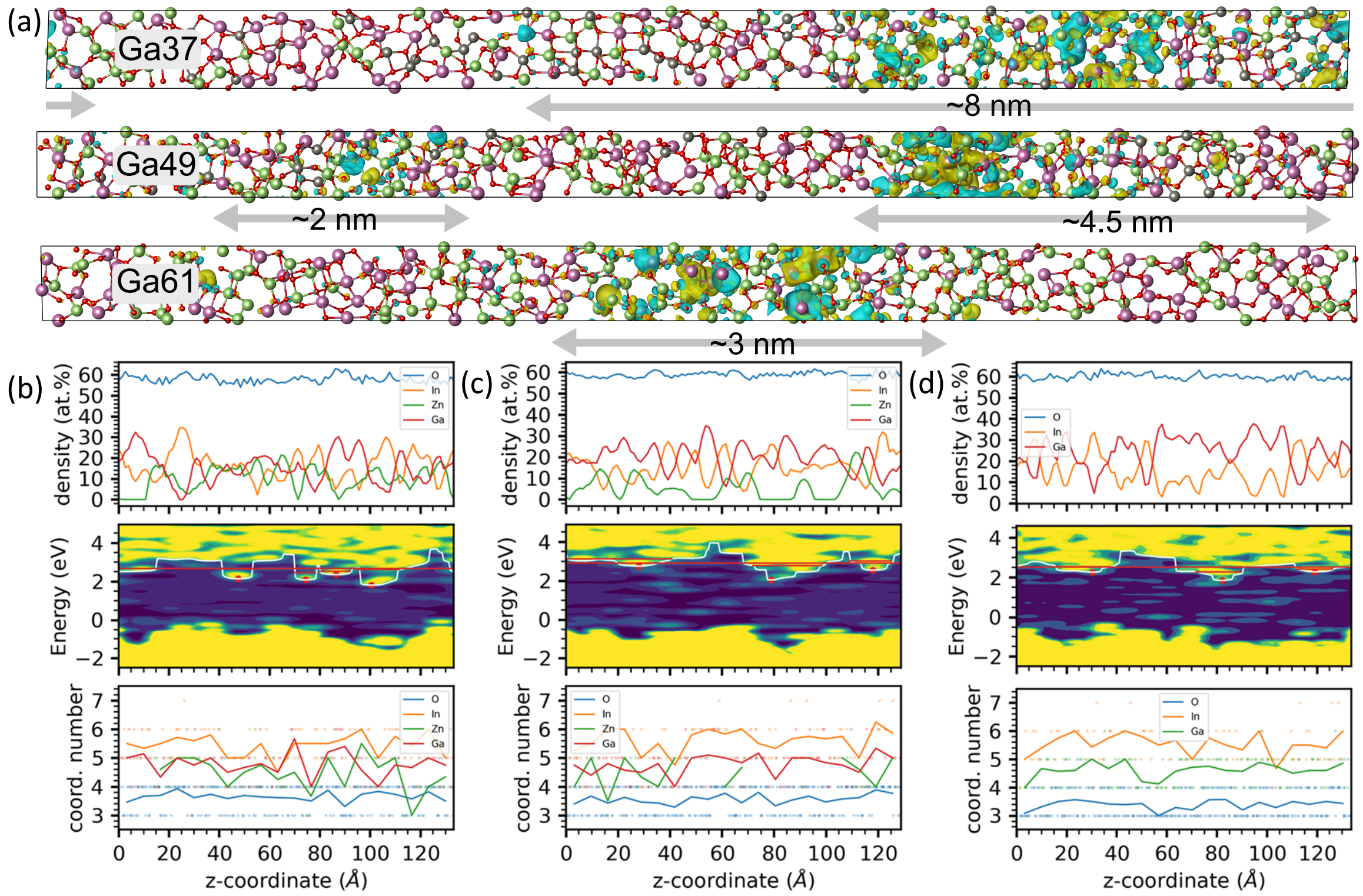

**Fig. 3 Electronic structure and structural analysis of three selected models in first-principles computations.** (a) Examples of conduction orbitals and the atomistic structure, oxygen in red, indium in pink, gallium in green and zinc in grey. (b)-(d) The elemental density, local density of states, and average elemental coordination for the Ga37 (b), Ga49 (c), and Ga61 (d) composition. The local density of states is averaged over the two short dimensions. The white contoured lines are the fixed threshold for tracing the local energy minimum. A direct one-to-one relation with average coordination numbers or with the atomic percentages of the metal species is not distinguishable in this computation.

## Energy landscape and electron transport model

The experimental transport characteristics and first-principles calculations consistently indicate that conduction states in a-IGZO are neither fully delocalized nor purely localized. Instead, electrons retain coherence only over finite spatial regions, cannot achieve fully degenerate metallic conduction over their entire transport path. This leads to a heterogeneous energy landscape characterized by the formation of potential wells (Fig. 4a). When electron states below the $E_F$ are occupied, the potential wells define conducting regions where electrons maintain coherence and propagate as quasi-free plane waves over a limited range. Outside these regions, stronger energetic and structural disorder gives rise to insulating barriers that interrupt global coherence. The initial Fermi level $E_{F0}$, set by composition, doping, metal work function, etc., defines the initial size of conducting and insulating regions. With increasing $E_F$ (e.g. under gate bias), the depth and spatial extent of the occupied potential wells increase, causing the conducting regions to expand and electron wavefunctions to spread further in space. At sufficiently high $E_F$, the spatial overlap between neighboring wave packets becomes significant, and the system approaches the limiting case of a fully delocalized band, whose edge we denote as $E_{dloc}$. Under typical operating conditions of a-IGZO transistors, however, transport occurs predominantly in the intermediate regime, where coherence remains local but global degeneracy is not achieved.

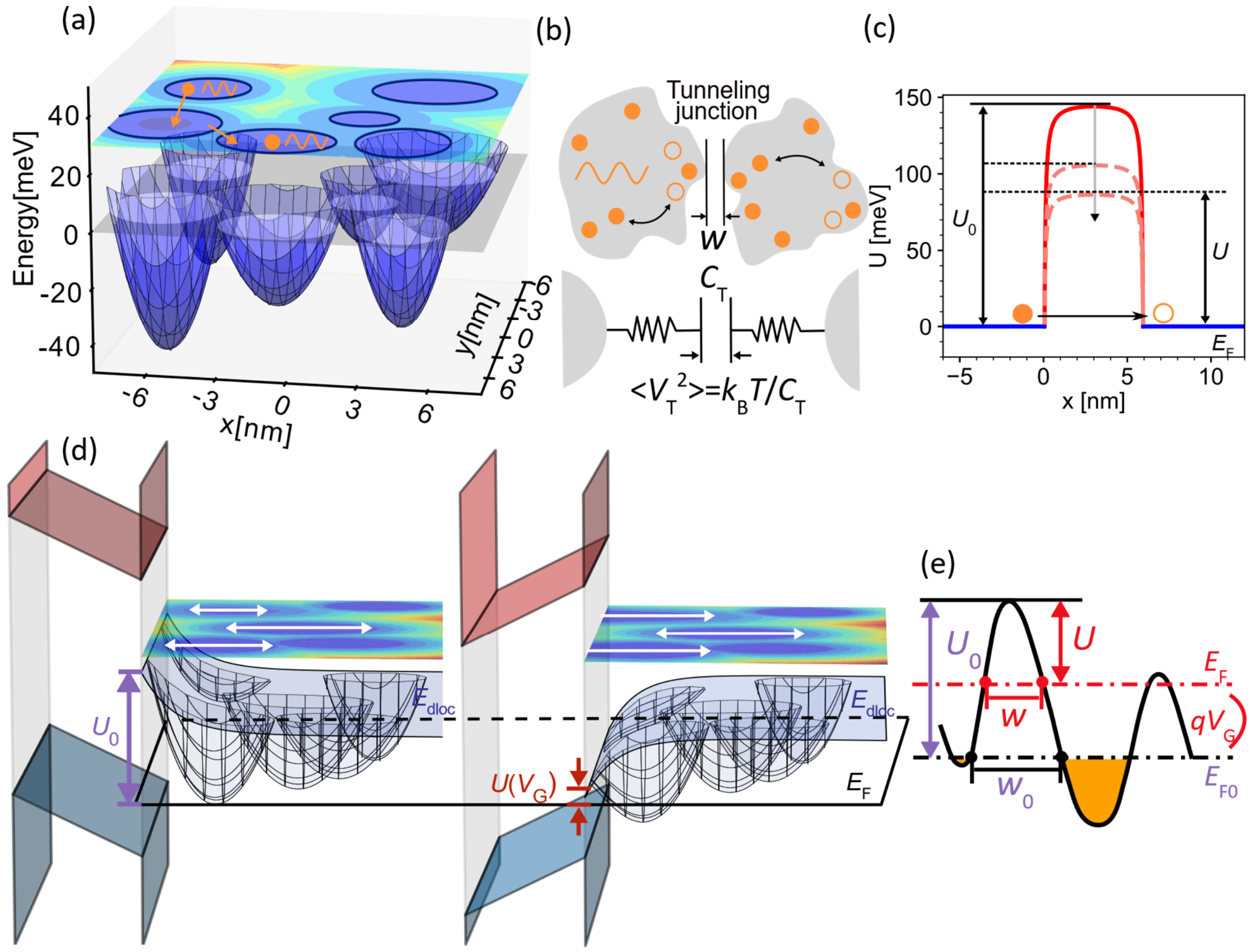

**Fig. 4 New heterogeneous energy landscape and FEAFIT transport model of a-IGZO** (a) Schematic of the proposed potential-well energy landscape. The gray plane at $E = 0$ eV represents fermi level. Occupied states (blue regions) consist of conducting regions, outlined by black solid circles in the top-view contour plot. Inside these regions, electrons behave as quasi-free plane waves (orange). In contrast, the surrounding insulating regions form energy barriers that hinder electron transport. (b) Schematic of tunnel junction between two conducting regions in FEAFIT model. The tunnel junction with width $w$ is modelled as a parallel-plate capacitor $C_T$. Owing to the random motions of electrons, there are excess or deficit of charges on the tunnel junction surfaces. (c) Simulated reduction of the tunnelling barrier due to temperature-induced voltage fluctuations in our model. (d) Band diagram with energy potential wells. Top-view contour plot is cut at $E_F$. Left: without applied $V_G$, $E_{dloc}$ bending upward, initial barrier $U_0$ labelled as purple is higher and conducting regions marked by white arrows are smaller. Right: under positive $V_G$, $E_{dloc}$ bending downward, enlarging the conducting regions and lowering the barrier height $U(V_G)$. (e) Similar to (d), but with fixed energy profile and a tunable $E_F$. The downward band bending corresponds to an upward shift in $E_F$, also reducing the junction width from $w_0$ to $w(V_G)$.

In such a landscape, we argue that long-range transport is governed by tunnelling through insulating gaps between the conducting regions. In a-IGZO, the gaps are expected to be nanoscale, because of the high extracted density of tail states and large spread range of electron wavepackets from DFT computations. Each gap can be regarded as a small capacitor (see Fig. 4b), whose capacitance $C_T = \varepsilon A/w$, with area $A$ and separated width $w$. By the equipartition theorem, the mean-square voltage fluctuation across the gap is given by $\langle V_T^2 \rangle = k_B T / C_T$. Because the capacitance is small for nanoscale junctions, these thermal voltage fluctuations are large and strongly modulate the barrier height and width[42]. Fig. 4c shows that the simulated tunnelling barrier is reduced by the thermally activated voltage fluctuation. We describe this process using a field-effect-aware fluctuation-induced tunnelling (FEAFIT) model. Within this model, the conductivity can be derived as

$$\sigma = \sigma_0 \exp\left[-\frac{T_1(V_G)}{T} S_1(T) - \frac{T_1}{T_0}(V_G) S_0(T)\right], \qquad (1)$$

where $T_1$ sets the activation scale and reflects the characteristic barrier energy linked to disorder, while the ratio $T_1/T_0$ determines the saturation value in the low-temperature limit, giving the combined effect of barrier height and width on elastic tunnelling. Here $S_1(T) = (\eta^*)^2$, $S_0(T) = \psi(\eta^*)$, with $\eta^*(T)$ the most-probable fluctuation in the thermal averaging. $S_1(T)$ rises from 0 at low temperature to 1 at high temperature, while $S_0(T)$ falls from 1 to 0. Together they describe the smooth crossover from elastic tunnelling to Arrhenius-like activated transport with slope $T_1$. This formulation assumes negligible charging energy in the conducting regions, validity of the WKB approximation[43], and that detailed barrier shape (image-force corrected rectangular or parabolic) only affects the tunnelling action and is absorbed into the effective parameters $T_1$ and $T_1/T_0$. Note S5 provides explicit description and the derivation of the model.

The influence of the gate field enters naturally through its effect on the tunnelling barrier. At zero gate bias, the tunnelling junction is characterized by an initial barrier height $U_0$ and width $w_0$, both controlled by the position of $E_{F0}$, which governs the formation of the conducting regions and the insulating gaps between them. As illustrated in Fig. 4d − e, applying a positive gate bias bends the bands downward at the channel surface. This reduces the potential difference between $E_{dloc}$ and the $E_F$, leading to a smaller effective barrier height $U(V_G)$. At the same time, the spatial extent of occupied potential wells increases, giving a reduced barrier width $w(V_G)$. In the FEAFIT framework, these changes directly map into the transport parameters. A reduction of $U$ decreases the activation scale $T_1(V_G)$, since $T_1$ reflects the barrier energy. Simultaneous reduction of both $U$ and $w$ decreases the ratio $T_1/T_0(V_G)$ (Eq. S25 and S28 in Supplementary Note S5). Thus, gate bias enhances tunnelling conductivity by lowering the barrier height and narrowing its width, which is captured by the gate dependence of $T_1(V_G)$ and $T_1/T_0(V_G)$ in FEAFIT.

**Model validation and composition-dependent evolution of transport parameters**

As shown in Figs. 5a – c, the FEAFIT model reproduces the experimental conductivity of all three a-IGZO compositions over the full temperature range (10 – 300 K) and for different gate voltages. Without dividing the data into separate regimes, the unified framework accounts for four key features: Arrhenius activation at high temperature, the gradual crossover to temperature-independent behaviour at low-temperature, the suppression of this plateau in Ga-rich compositions, and the overall enhancement of conductivity with increasing $V_G$.

The extracted transport parameters follow the expected trends. At high temperature the conductivity reduces to $\sigma \propto \exp[-T_1/T]$, where $T_1$ reflects the barrier energy scale. As Ga concentration increases, the higher energetic disorder leads to larger activation energies (Fig. 2b) and correspondingly larger $T_1$ values (Fig. 5d). At very low temperature the transport reduces to $\sigma \propto \exp[-T_1/T_0]$, and the ratio $T_1/T_0$ sets the tunnelling plateau and relates to tunneling barrier width and height. This ratio increases with Ga content, consistent with stronger disorder producing wider and higher barriers (Fig. 5e). In Ga-rich samples, however, this plateau is suppressed. The reduced size of conducting regions and the higher barriers shift the mechanism away from tunnelling, leading to a renewed temperature dependence of conductivity. This behaviour explains the finite value of activation energy in third regime, $E_{a,3}$, and the existent low temperature dependence of $\sigma$ in Ga49 and Ga61 (Fig. 5b–c). Furthermore, gate bias decreases both $T_1$ and $T_1/T_0$ in all compositions, in agreement with the predicted reduction of barrier height and width under positive $V_G$.

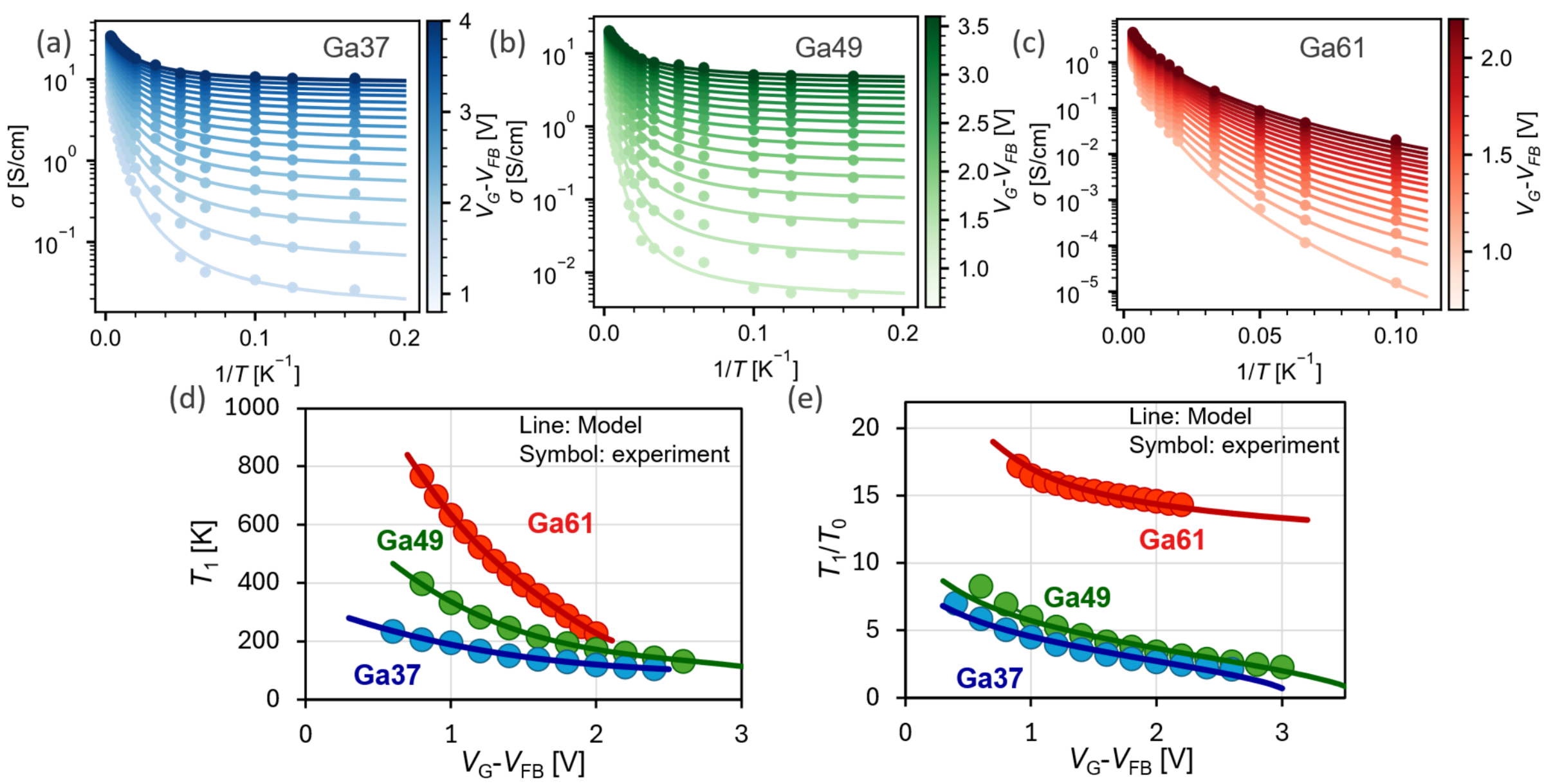

**Fig. 5 Model validation and correlation of transport parameters.** Comparison between experimentally extracted conductivity ($V_D$ = 0.05 V) and the conductivity calculated using the FEAFIT transport model for Ga37 (a), Ga49 (b), and Ga61 (c) IGZO. (d–e) The gate bias and composition dependence of model parameters. Symbols represent the experimentally extracted $T_1$ (d) and $T_1/T_0$ (e), while solid lines correspond to the values from our transport model, using the physical parameters $w_0$ and $A$ extracted from fitting.

Beyond reproducing conductivity trends, the model also allows extraction of the initial width $w_0$ and the effective area $A$, defined at zero bias. Here, $w_0$ is the mean separation between conducting regions, and $A$ is the average area of the effective capacitor. $w_0$ increases systematically with Ga content, from ~4 nm in Ga37 to ~8 nm in Ga61, while $A$ rises from ~$10^{-16}$ $m^2$ to more than $10^{-15}$ $m^2$, as shown in Fig. 6a-b. In parallel, DFT calculations also show that the separation between conduction pockets increases with Ga content and decrease with energy separation between $E_{dloc}$ and $E_F$ (Fig. 6c), in line with the extracted $w_0$ from FEAFIT (Fig. 6d). Over the same composition range, monotonic anticorrelation between $w_0$ and $\sigma$ (from 20 K to 300 K) in Fig. 6a indicates that $w_0$ can serve as a physical descriptor linking atomic composition to macroscopic conduction in a-IGZO.

This agreement between FEAFIT transport model and DFT validates the proposed potential-well landscape. Beyond describing the temperature and gate-voltage dependence of conductivity, the FEAFIT framework enables extraction of barrier parameters with clear physical interpretation. The composition trend is clear: higher Ga content produces larger insulating gaps $w$ and higher barriers, corresponding to smaller conducting regions and a Fermi level further from the fully delocalized band edge, which is consistent with the observed reduction in conductivity.

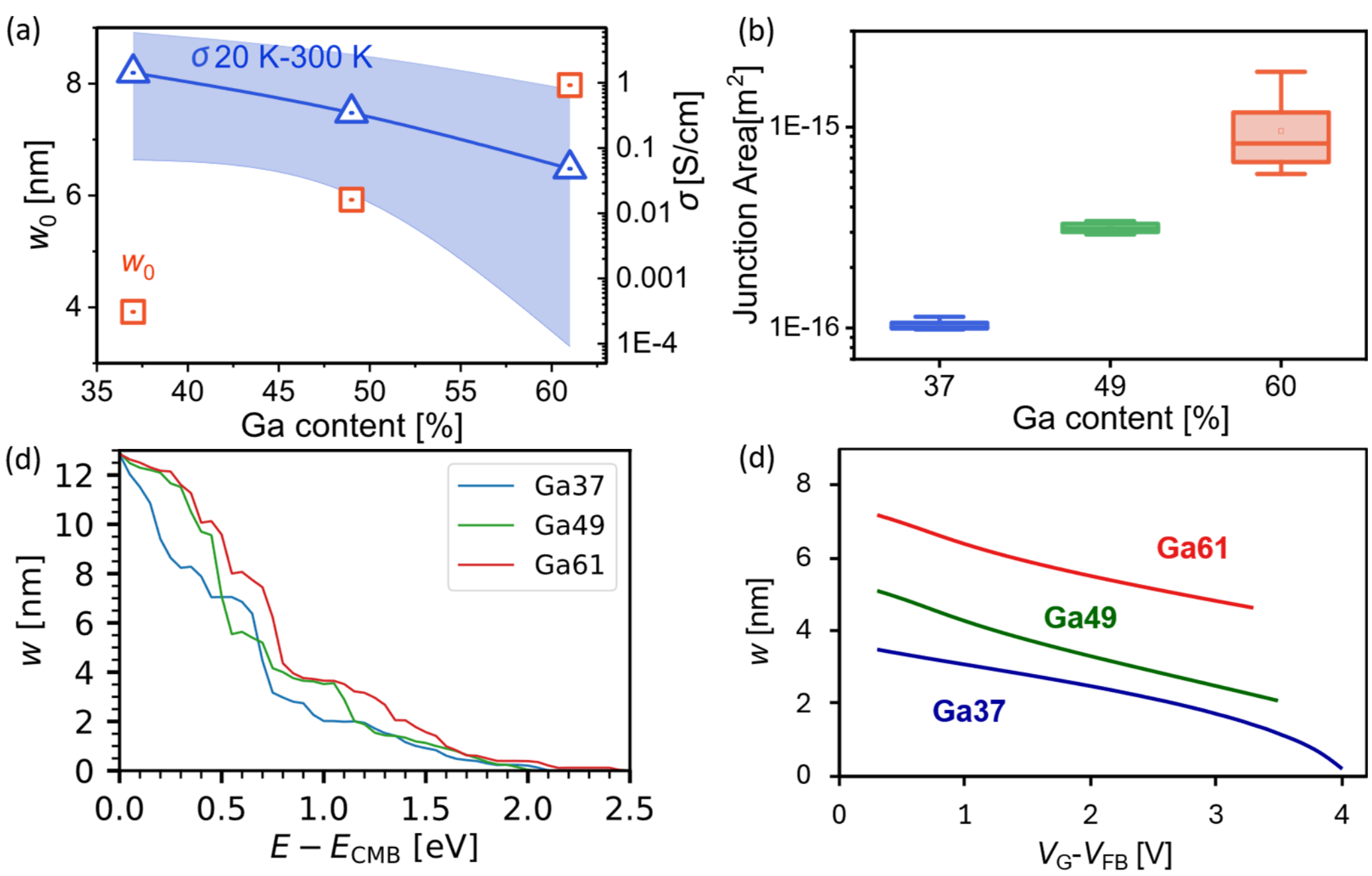

**Fig. 6 Tunneling barrier parameters.** (a) Composition dependence of initial tunnelling barrier width $w_0$ extracted from FEAFIT (left axis). The shaded region (right axis) represents the experimentally measured conductivity $\sigma$ range between 20 K and 300 K at fixed $V_G - V_{FB} = 0.4\ V$ and $V_D = 0.05V$. The triangles are the conductivity of 100 K. (b) Composition dependence of the average area *A* of the effective capacitor, extracted from FEAFIT. (c) First principles computation results of barrier width *w* decreasing with larger $E_F$. $E_{CMB}$ is the absolute minimum of the globally continue electron states. The larger the energy separation $E$- $E_{CMB}$, the closer $E_F$ lies the to fully continuous band edge $E_{dloc}$. *w* of Ga-rich IGZO is larger in the energy range that conduction mostly happens. (d) The extracted *w* of three compositions from FEAFIT decrease with the applied gate voltage. Larger gate voltages shift $E_F$ closer to$E_{dloc}$, resulting in a reduced barrier width *w*.

## Discussion

Our proposed view of the Fermi surface morphology and the associated transport mechanism fundamentally differ from both Kamiya's percolation model[16] and classical hopping theories[23,25]. In Kamiya's framework, electron states are assumed to fill into a continuous Fermi surface, enabling electrons to percolate coherently across the potential landscape, even in the presence of high barriers. In contrast, classical hopping models such as Multiple Trapping and Release (MTR) and Variable Range Hopping (VRH) describe carriers as being confined to individual atomic sites, requiring phonon-assisted jumps to move between them.

Our findings indicate a different scenario: the Fermi level does not extend into the fully delocalized conduction band but remains within a dense tail of states below the edge of fully delocalized states. At high filling levels, these states form finite-size "puddles" of coherent electrons separated by insulating gaps, resembling a dry lake bed picture. These puddles cannot merge into a continuous Fermi surface, preventing global coherence. At lower filling levels (low $E_F$), conduction becomes even more fragmented, but still not limited to single-site hopping. Instead, transport is best described as fluctuation-assisted tunnelling between these coherent regions. Rather than invoking percolation at high temperature and hopping at low temperature, the FEAFIT framework provides a unified description based on two key factors: the filling level and the degree of coherence within tail states.

Experimental observations and first-principles calculations strongly support this picture. Temperature-dependent transport measurements in Ga37, Ga49, and Ga61 reveal composition-dependent disorder and tail broadening, while field-effect Hall measurements and DFT pocket analyses confirm the partial coherence of electronic states. The FEAFIT model consistently reproduces conductivity across all temperatures and gate voltages, and even aligns with previously reported data (Fig. S22). Most importantly, it yields barrier dimensionality parameters that agree in both magnitude and composition dependence with DFT predictions. This is the first time such physically meaningful tunneling parameters have been extracted for a-IGZO. Collectively, these results advance the fundamental understanding of charge transport in amorphous oxides and establish a physically grounded framework for material co-design and device performance optimization.

## Methods

**Different composition a-IGZO wafers fabrication.** Back-gated a-IGZO TFTs with gate length and width of 10 μm are used. The devices are processed on a Silicon (p++) substrate. After gate-dielectrics deposition, a 12nm thick a-IGZO films are deposited by co-sputtering three different monoxide targets ($InO_x$, $GaO_x$, $ZnO_x$) in a combinatorial fashion, allowing to tune any a-IGZO composition. A 100 nm PECVD SiO2 encapsulation layer is deposited. PVD TiN liner and a CVD W filling metals are used for the S/D contacts. Finally, a 350°C $O_2$ annealing step for 1 hour is applied.

**Temperature dependent IV measurement.** Temperature-dependent electrical measurements were conducted using a Keithley 2636B sourcemeter in combination with a Lakeshore CPX-LVT cryostat. After dicing the full wafers into small pieces, the samples were mounted in a vacuum chamber pumped to 1e-3 Pa for cryogenic measurement. Wire bonding to a printed circuit board provided stable electrical contacts and minimized additional contact resistance at cryogenic temperatures. Prior to each measurement, thermal equilibrium was ensured by holding the device at the target temperature for 15 minutes. During transport measurements, the drain voltage ($V_D$) was held at 50 mV, and the common back gate was swept to acquire the transfer IV characteristics.

**Ellipsometry measurement.** Optical thickness measurements were executed on a-IGZO films with an average thickness range of 24 – 50 nm, deposited on Si using spectroscopic ellipsometry. A KLA SCD100 tool handling 300 mm Si wafers was used with an optical wavelength from 300 – 800 nm. Refractive index and thickness were fitted using the Harmonic Oscillator model. The Tauc Plot method was employed to determine the optical band gap of the film. Here, the attenuation was extracted from the imaginary part of the complex refractive index and its square root is plotted as function of energy[44]. The bandgap is than obtained from the linear fit from 4 eV down to lower values.

**Hall measurement.** To perform the Hall test, samples ($W \times L = 6 \times 35$ um$^2$, fabrication see Note S3) were placed into a Lakeshore CPX-LVT cryostat, pumped to 1e-3 Pa and cooled. The Hall tests were conducted from 100 K to 300 K. In each measurement, magnetic field $B$ was swap from -2.1 T to 1.5 T in steps of 0.1 T. For electrical measurements, device source was biased at 0 V, drain at 50 mV DC + 5 mV AC (11 Hz), and the gate bias $V_G$ applied through the backside of the sample was swept from -1 to 4 V. Source to drain current $I_{XX}$ was measured, as well as the potential $V_{XX}$ between the two inner pads (#4 and #5), together with Hall voltage $V_{Hall}$, measured between two transverse pads (#3 and #4). The Hall voltage $V_{Hall}$ was then plotted as a function of magnetic field, linearly fitted, allowing to extract the Hall coefficient $R_H = \frac{I_{XX} \cdot B}{V_{Hall}}$. Details see Note S4.

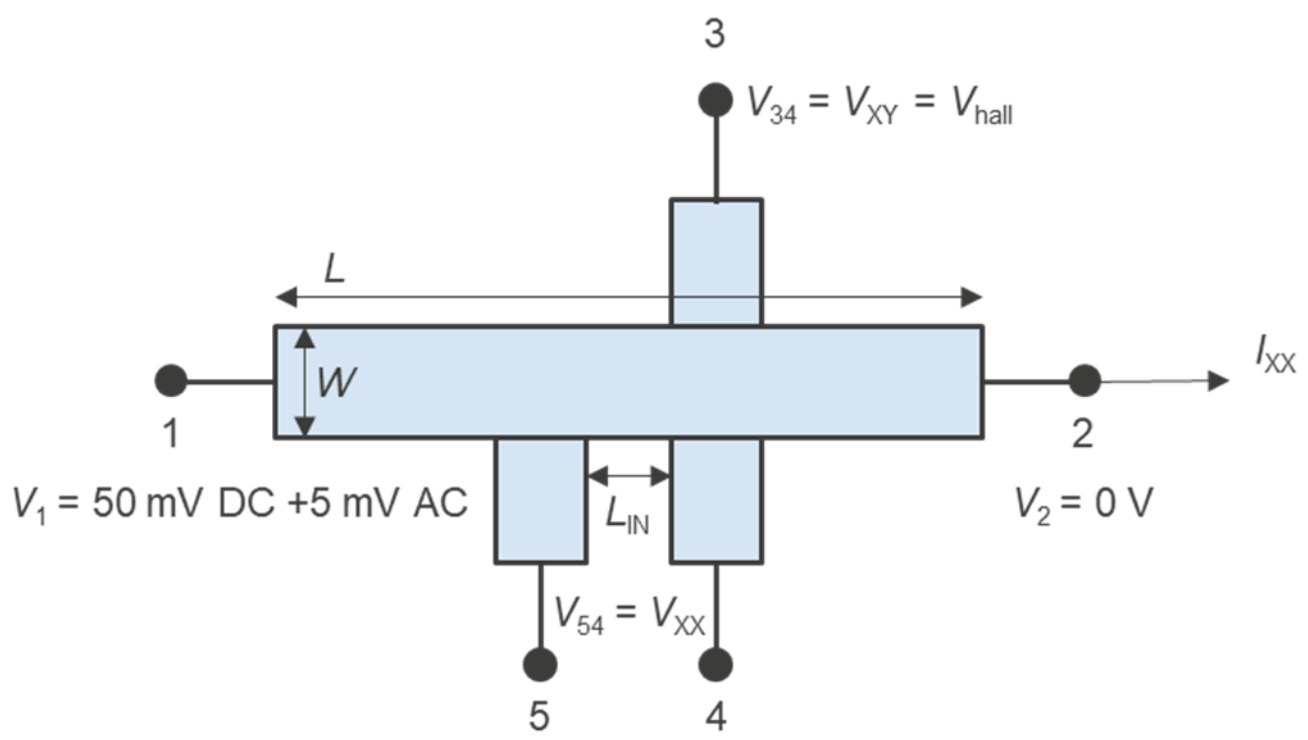

Fig. M1: Biasing schematic for Hall bar structure

**DFT.** First principles calculations using the density functional theory formalism have been employed using the CP2K software version 2023.2. The hybrid Gaussian and plane wave density functional scheme of CP2K[45–49] makes the dimensions of the systems needed the required cell dimension computationally feasible. We used the PBEsol generalized gradient approximation for the exchange-correlation functional[50,51]. The standard double-ζ valence plus polarization (DZVP) basis sets[52] and pseudo potentials[53–55] provided with CP2K were used. All calculations were performed using a single **k**-point (Γ) to prevent effects caused by the artificial periodicity introduced by the supercell approach from influencing the results. For the structure optimization, we used a maximum geometry change convergence criterion of 5 mBohr and a force convergence criterion of 1 mEH/Bohr. We used a target accuracy for the electronic self-consistency convergence of $1 \times 10^{-6}$ $E_{\mathrm{H}}$. The preparation, execution, monitoring, and postprocessing of all computations reported in this work have been facilitated by our in-house Python package. The amorphous structural models are generated using the decorate-and-relax method proposed by Drabold et al.[56]. In our experience, this approach leads to less defective structures at lower computational costs than melt-and-quench methodologies[41,57]. The structural optimization, the 'relax' part, uses a combination of the Broyden−Fletcher−Goldfarb−Shanno (BFGS) algorithm[58–61] and time-stamped force-bias MonteCarlo (TFMC)[62,63].

## Acknowledgement

This work has been enabled in part by the NanoIC pilot line. The acquisition and operation are jointly funded by the Chips Joint Undertaking, through the European Union's Digital Europe (101183266) and Horizon Europe programs (101183277), as well as by the participating states Belgium (Flanders), France, Germany, Finland, Ireland and Romania. For more information, visit nanoic-project.eu. The first author deeply thanks Jiawei Wang for intellectual guidance and helpful discussions during the doctoral studies, which contributed to the first author's understanding of charge transport in IGZO.

# Supplementary Information

## Revealing the Partial Coherent Nature of Transport in IGZO

Ying Zhao[1*], Michiel J. van Setten[1], Anastasiia Kruv[1], Pietro Rinaudo[1,3], Harold Dekkers[1], Jacopo Franco[1], Ben Kaczer[1], Mischa Thesberg[2], Gerhard Rzepa[2], Franz Schanovsky[2], Attilio Belmonte[1], Gouri Sankar Kar[1], Adrian Chasin[1]

[1]imec, Leuven, Belgium
[2]Global TCAD Solutions (GTS) GmbH, Vienna, Austria
[3]KULeuven, Leuven, Belgium

*Email: ying.zhao@imec.com

## Part. 1 Film calibration and basic electronic properties of a-IGZO transistors

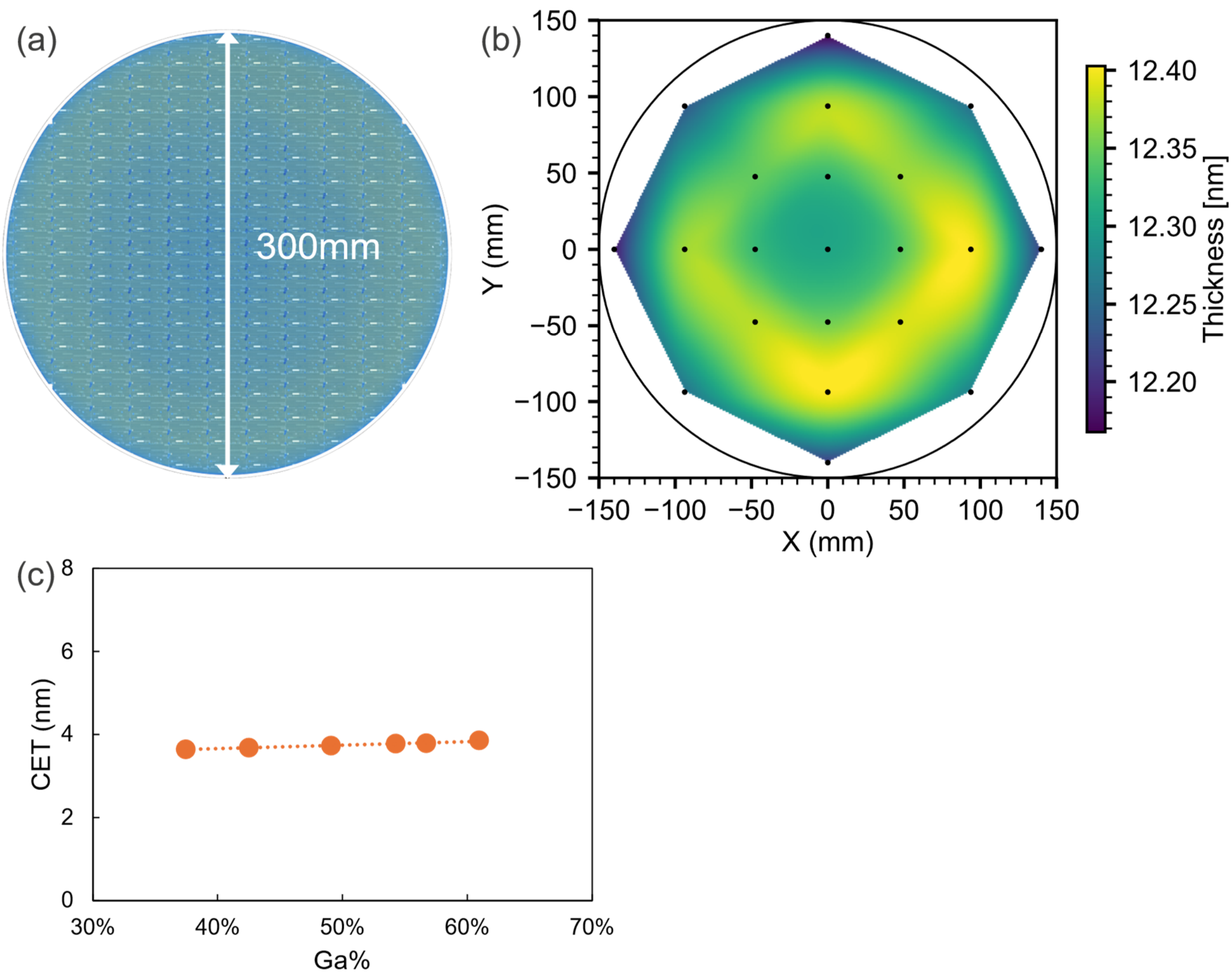

**Fig. S1** (a) Photograph of the fabricated 300 mm-wafer of a-IGZO. (b) The deposited a-IGZO thickness map, the standard deviation *s* = 0.08 nm. (b) The relation between Capacitance Equivalent Thickness (CET) of a-IGZO TFTs and Ga concentration of a-IGZO film. CET is obtained from the measured capacitance: CET = $\varepsilon_{SiO_2} \times area/C_{meas}$.

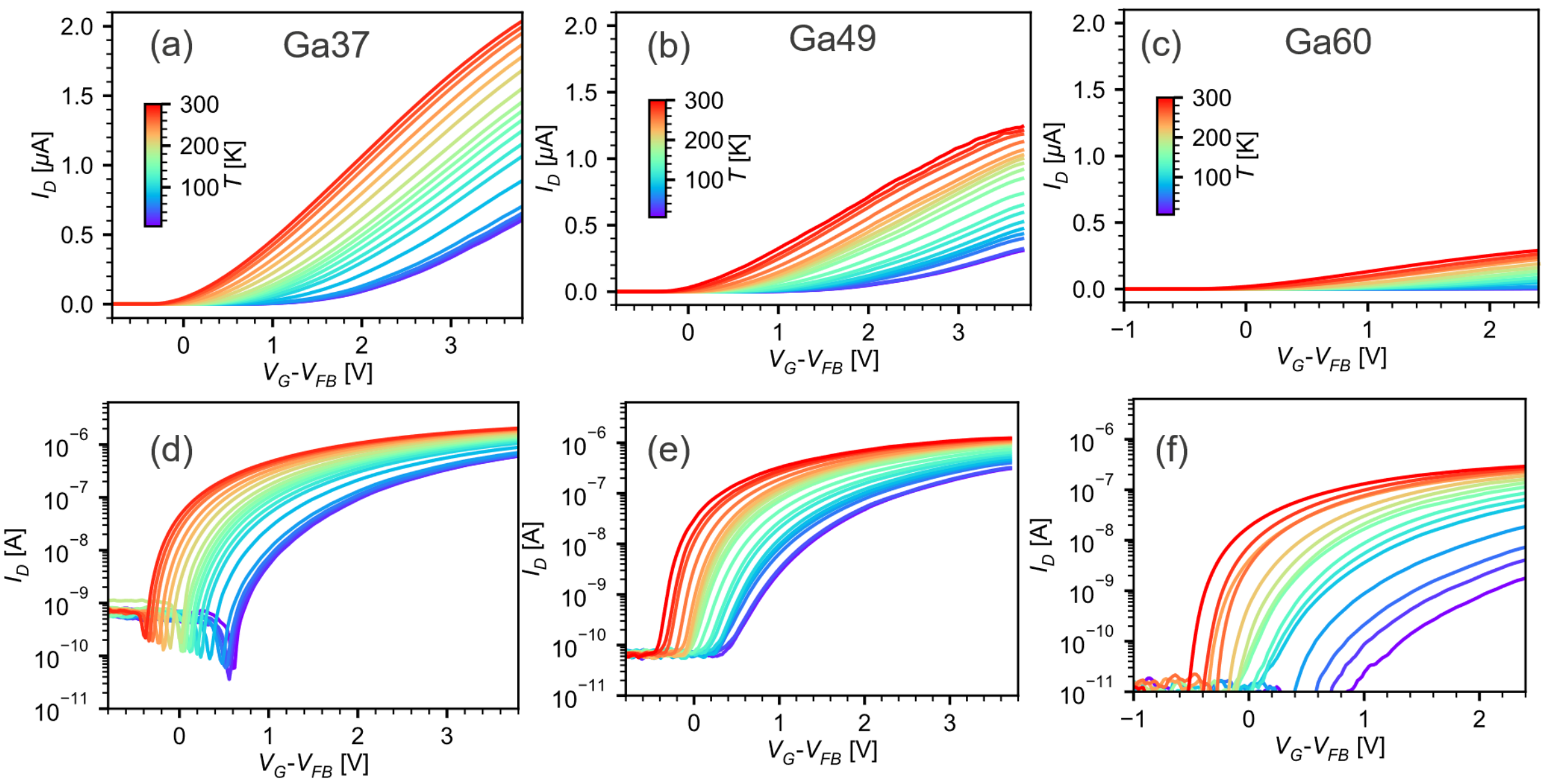

**Fig. S2** The transfer characteristics of a-IGZO TFTs with three compositions in linear scale (a)-(c) and logarithm scale (d)-(f), from cryogenic temperature to room temperature.

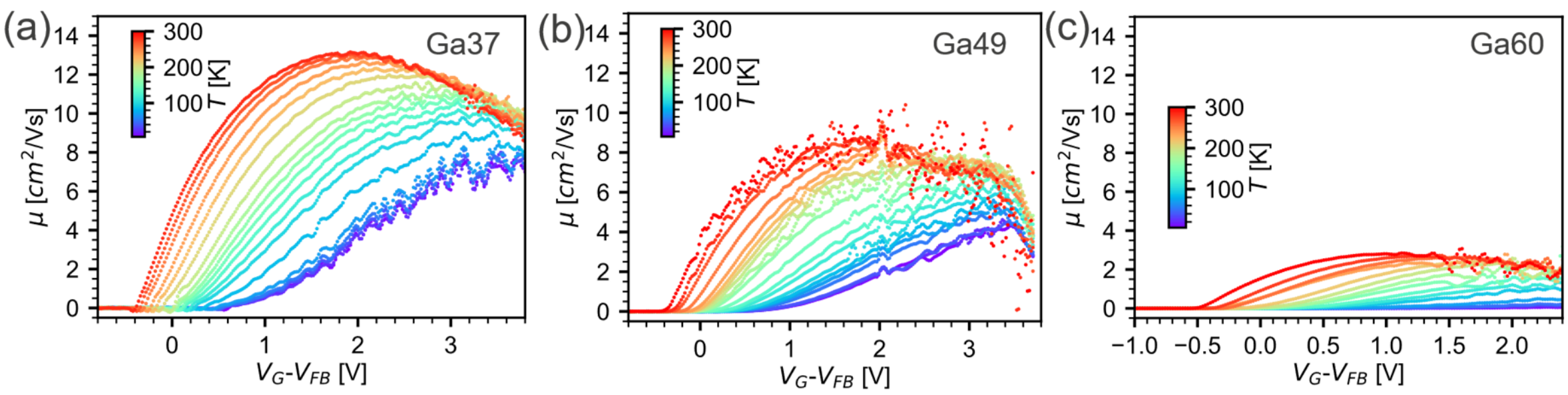

**Fig. S3** The extracted mobility from cryogenic temperature to room temperature of three a-IGZO compositions. $\mu = \frac{L}{WV_D C_i}\left(\frac{\partial I_D}{\partial V_G}\right)$.

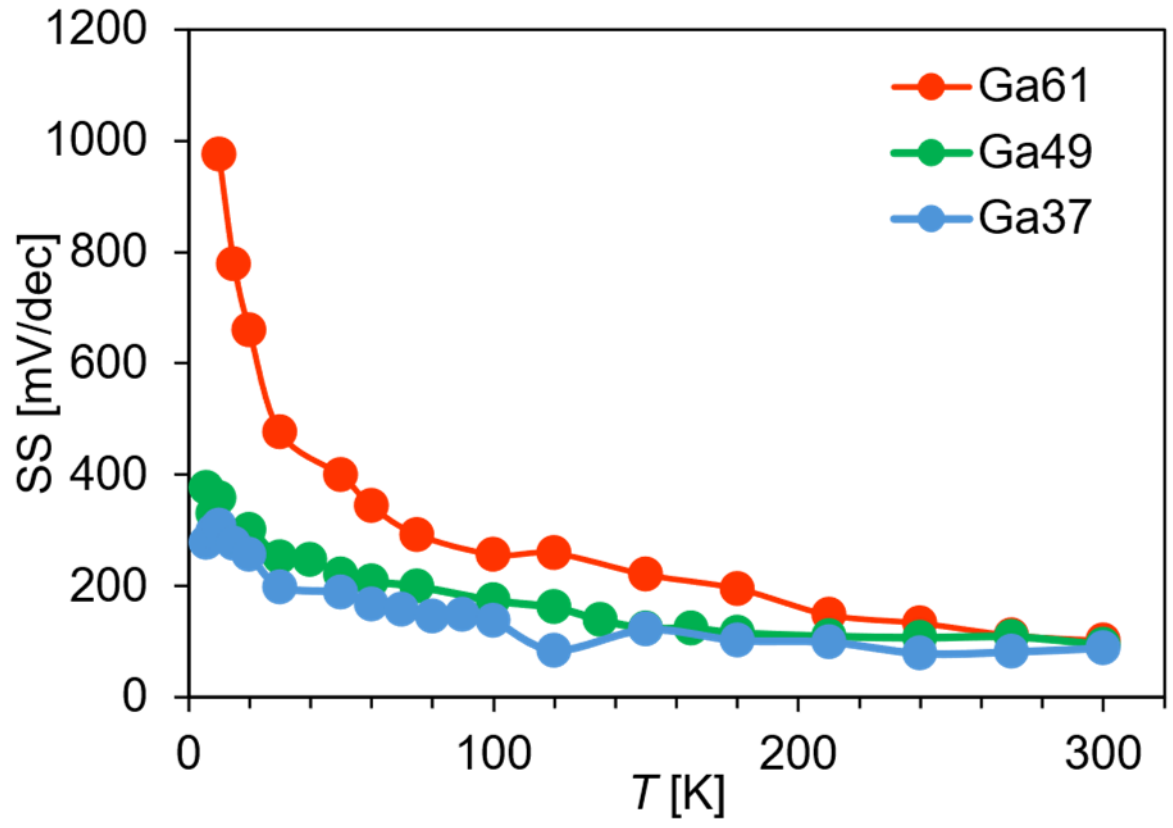

**Fig. S4** The extracted subthreshold swings of three a-IGZO compositions, $SS = \left(\frac{d(\log_{10} I_D)}{dV_G}\right)^{-1}$.

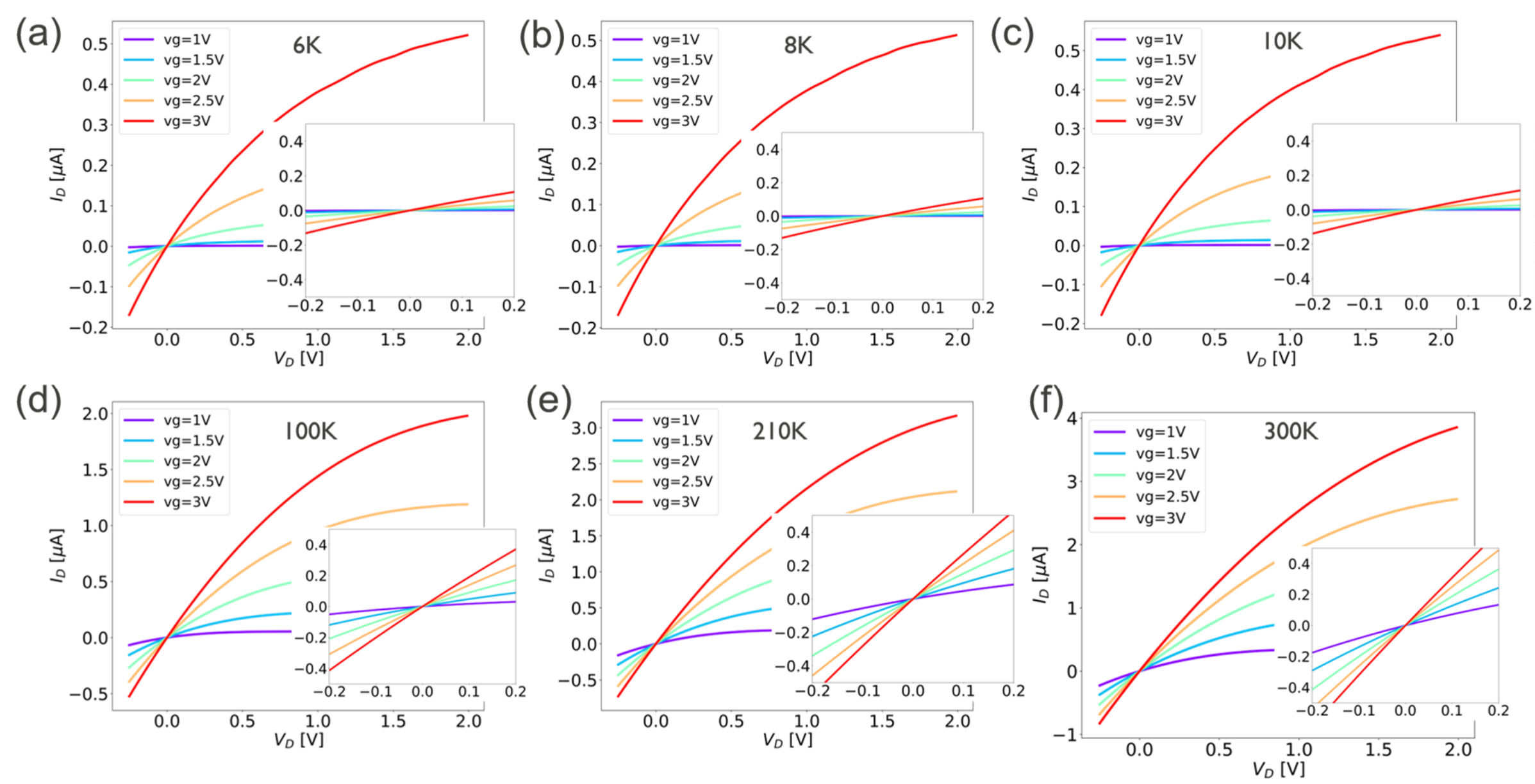

**Fig. S5** The output characteristics of Ga37 a-IGZO with various applied $V_G$ under 6 K (a), 8 K (b), 10 K (c), 100 K (d), 210 K (e), 300 K (f). Each inset shows a magnified view of the output curves near the origin.

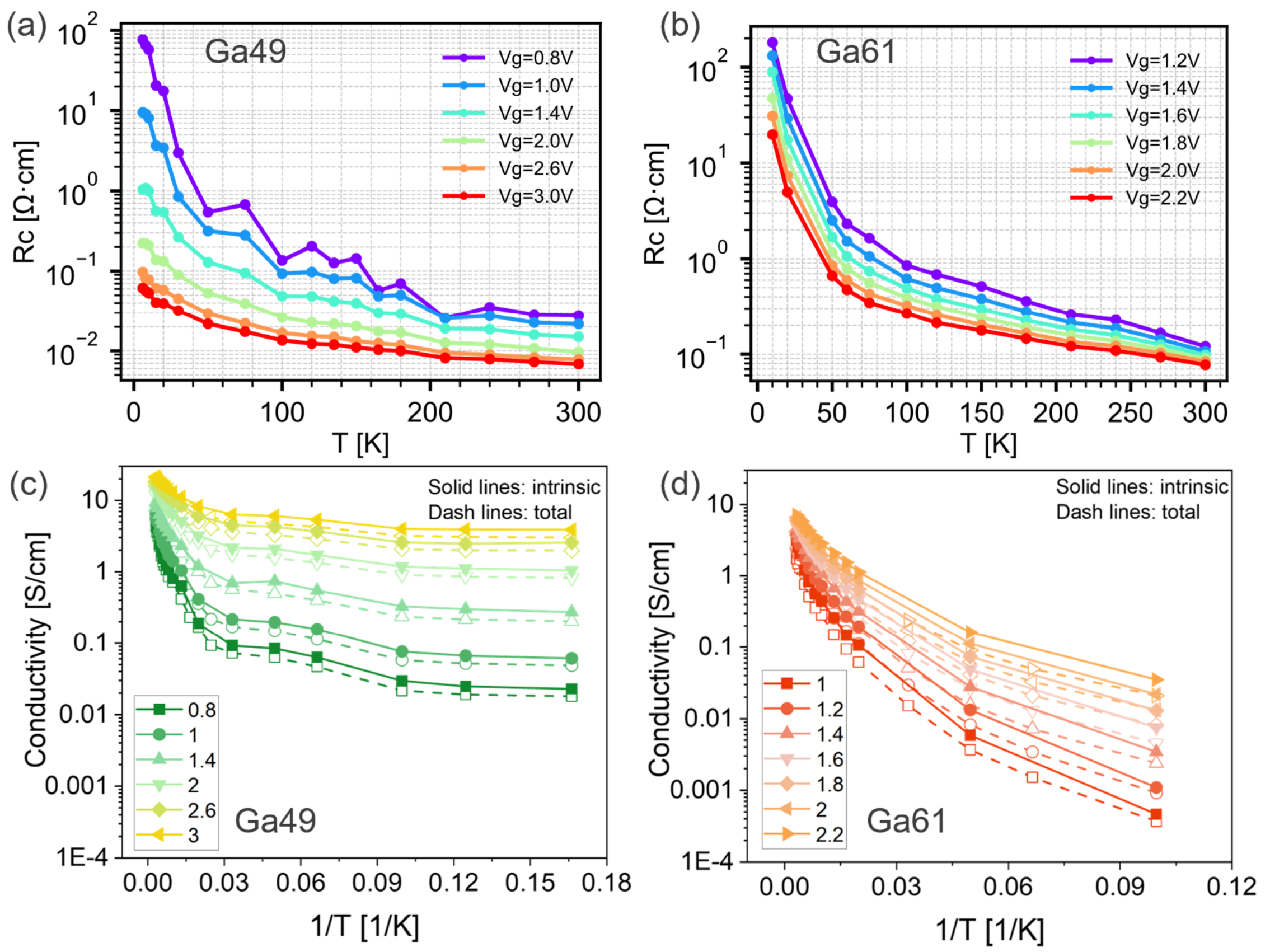

**Fig. S6** The contact resistance extracted by Transfer Line Method (TLM) of Ga49 (a) and Ga61 (b) a-IGZO. The intrinsic conductivity extracted by subtracting the obtained contact resistivity of Ga49 (c) and Ga61 (d) a-IGZO devices.

From the output characteristics of Ga37 a-IGZO in Fig. S5, the best conduction performance among three compositions, output curves maintain good linearity even at 6 K. It indicates that contact resistance is not a problem and limiting factor of Ga37 a-IGZO in the transport analysis. For poor conduction performance a-IGZO, Ga49 and Ga61, we extract the contact resistivity via Transfer line method (TLM) for different gate-length transistors, as shown in Fig. S6 (a) and (b). The resistivity of Ga49 and Ga61is small at room temperature and keep increasing with cooling down. This means the contact resistance is not dominated by Schottky barrier but the intrinsic disorder of a-IGZO film itself. We subtract total resistivity by the contact resistivity, in order to get the intrinsic conductivity of a-IGZO film. In Fig. S6 (c) and (d), the intrinsic conductivity only a little larger than the measured conductivity. The temperature, gate bias and composition dependence are all the same for intrinsic and measured conductivity. Even for Ga61 a-IGZO with inherently the poorest conductivity, contact resistance of our a-IGZO TFTs has a negligible impact on transport characteristics. Thus, we can conclude that all the transport characteristics we study and analyse on our a-IGZO TFTs are not impacted by contact resistance.

# Part 2 Temperature-dependent transport data and tail states characterization

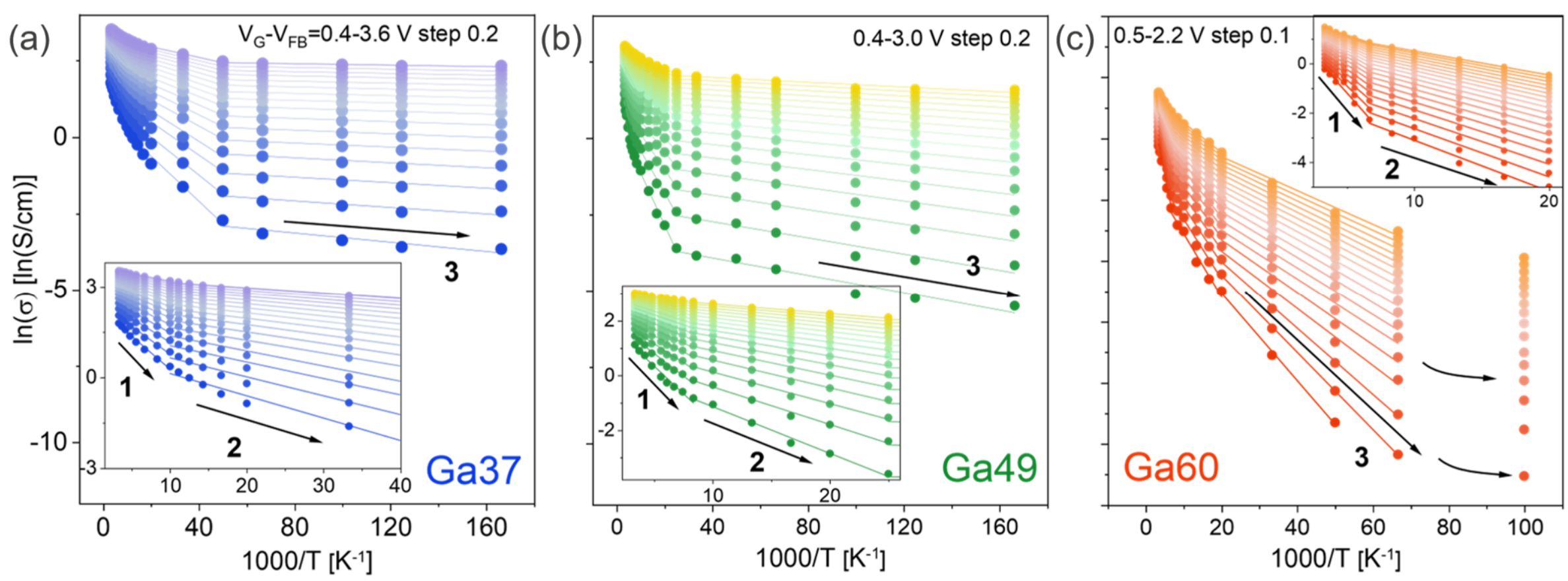

**Fig. S7** Temperature dependence of the extracted conductivity for three compositions. $\sigma$ is divided into three regimes for Arrhenius fitting ($\sigma \propto \exp\left(-E_{a,1,2,3}/k_B T\right)$, solid lines), labelled as regime 1-3. The inset provides a magnified view from 50 K to 300 K.

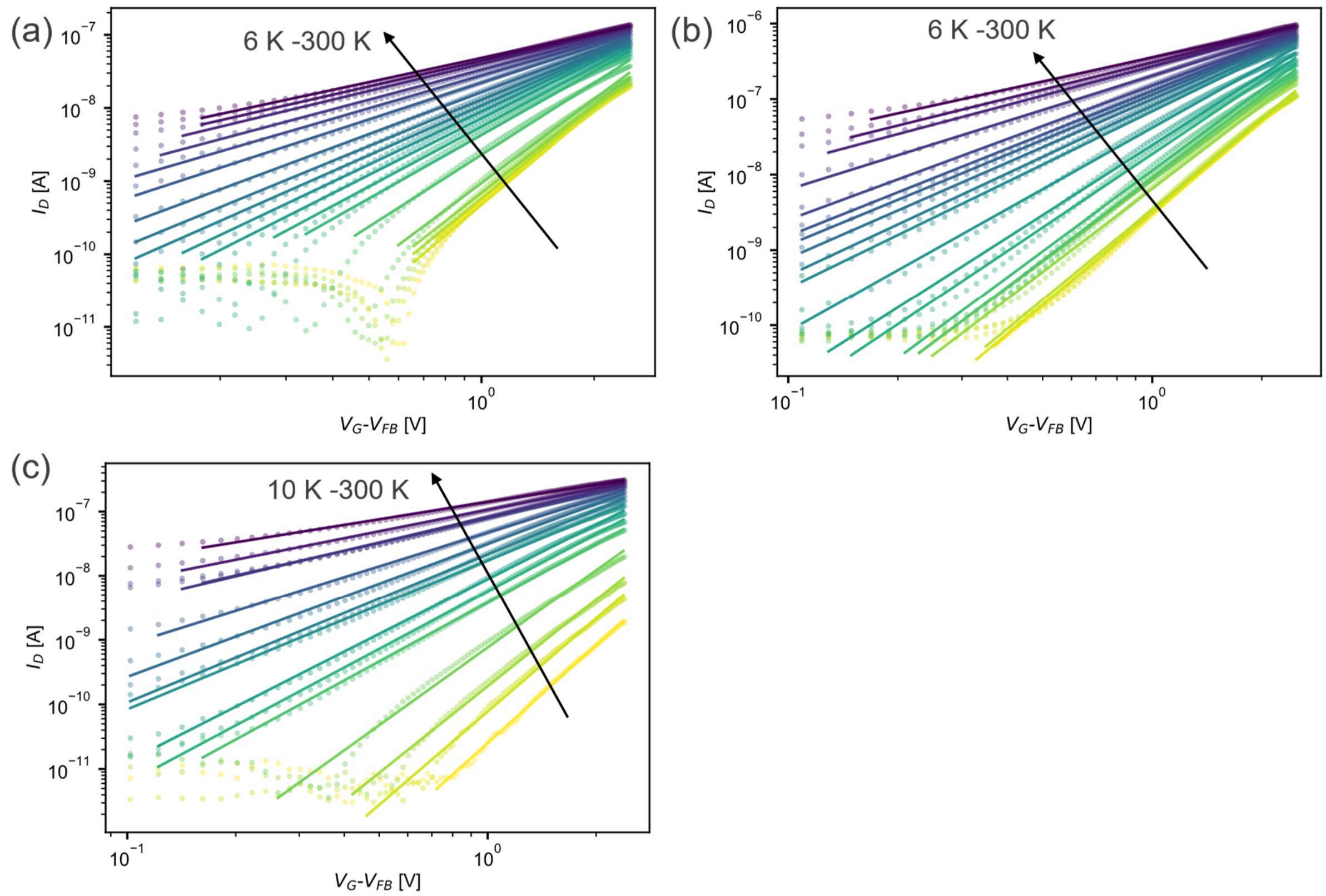

**Fig. S8** The power law dependency ($I_D \propto (V_G - V_{FB})^{\gamma}$) of conductivity of the three a-IGZO compositions Ga37 (a), Ga49 (b) and Ga61 (c), on the applied $V_G$-$V_{FB}$. The solid lines are the power law fitting, from which the exponent $\gamma$ can be extracted for further analysis [1].

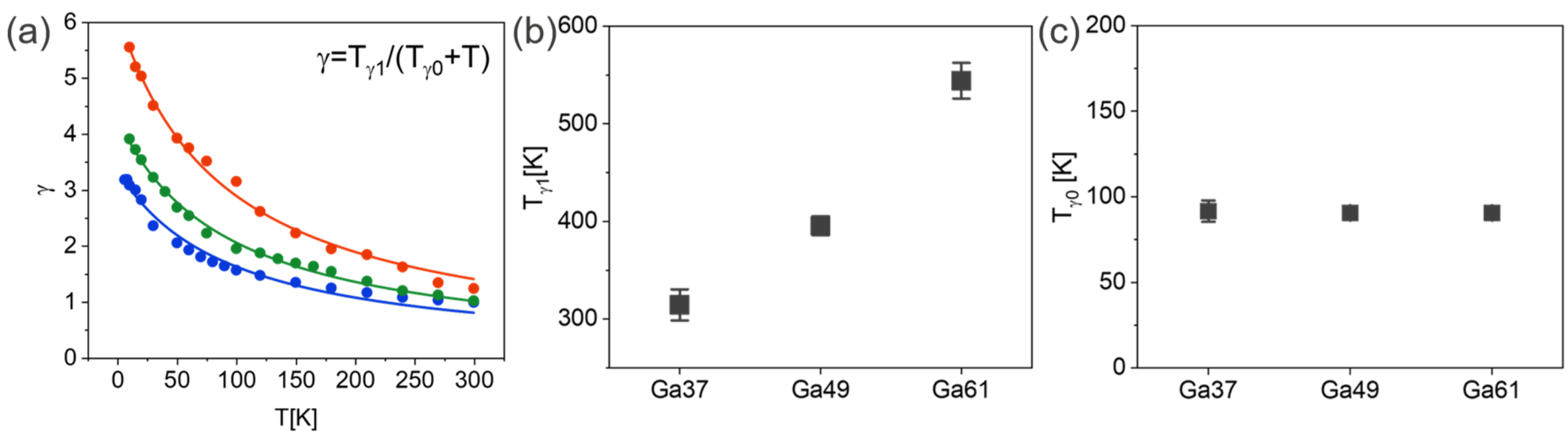

**Fig. S9** Power-law exponent $\gamma$ extracted from the fits (Fig. S8), plotted against temperature in linear scale (a) from 10 K to 300 K. The solid-color lines correspond to the FEAFIT model fitting ($\gamma = T_{\gamma 1}/(T_{\gamma 0} + T)$). The extracted $T_{\gamma 1}$ (b) and $T_{\gamma 0}$ (c) are from the FEAFIT model fitting.

## Note S1 Extraction of the trap density of states from experiment

The density of localized trap states in the sub-gap region was extracted from temperature-dependent transfer characteristics using an analytical method originally proposed by Lang[2]. The method relates the activation energy $E_a$, derived from the temperature dependence of field-effect conductivity $\sigma(T)$, to the gate-induced charge and hence to the energy-resolved trap density of states $N(E)$.

At each gate voltage $V_G$, the activation energy $E_a$ was extracted by fitting the Arrhenius plot of $\ln\sigma$ versus $1/T$ in the temperature range from 120 K to 300 K, the first regions of three compositions in Fig. S7. The resulting $E_a$ ($V_G$) dependence was then interpolated using a 5th-order polynomial to ensure smooth numerical derivatives.

Following Lang's approach, the trap DOS was computed from the first derivative of $E_a$ with respect to $V_G$ using the following expression

$$N(E) = \frac{C_i}{ea}\left(\frac{dE_a}{dV_G}\right)^{-1}, \quad \text{(S1)}$$

where $C_i = \varepsilon_0\varepsilon_i/t_i$is the areal gate capacitance (with equivalent dielectric permittivity $\varepsilon_i = \varepsilon_{SiO_2} = 3.9$, and $t_i$ = CET measured experimentally in Fig. S1, $e$ is the elementary charge, and $a$ is the effective thickness of the accumulation layer. In our analysis, $a$ was not treated as a fitting parameter but was instead extracted from TCAD simulations of the vertical charge density profile in accumulation respecting to the interface. The simulated charge distribution suggests that the carriers are confined within a few nanometers from the semiconductor/dielectric interface, and a value of $a$ increases slightly with Ga concentration, as shown in Table S1.

**Table S1 The extracted thickness of accumulation layer from TCAD**

| | Ga37 | Ga49 | Ga61 |
|---|---|---|---|
| $a$ [nm] | 4 | 4.5 | 5 |

This method allows for a direct estimation of *N(E)* in the relevant energy range. While absolute DOS values depend on the assumed accumulation thickness [2], our analysis focuses on the relative trends of energy disorder and trap state distribution among different material systems.

From the tail DOS extraction results, We also find that an exponential function $N_{\text{tail}}(E) = N_0\exp(-E/E_0)$ with a single slope $E_0$ cannot correctly fit the entire experimentally extracted tail states distribution. Instead, two exponential distributions are needed (Fig. 2c in main text): one with a narrow width ($E_{0,1}$) and high-density ($N_{0,1}$), and another with a broader width ($E_{0,2}$) and low-density ($N_{0,2}$). As the Ga concentration increases from 37% to 61%, $E_{0,1}$ changes from 6 meV to 12 meV and $E_{0,2}$ increases from 50 meV to 157 meV (inset of Fig. 2d in main text). The parameters of the tail states are calibrated by TCAD, confirming the existence of the two tails under $E_{\text{dloc}}$, shown in Note S2. The two tails of localized states represent different degrees of disorder and contribute to different transport behaviors. The first tail, characterized by a small $E_{0,1}$ and larger $N_{0,1}$, corresponds to localized electron states that are densely distributed near $E_{\text{dloc}}$. The second deeper tail distribution, with larger $E_{0,2}$ and smaller $N_{0,2}$, indicates a greater degree of disorder and represents moderate localized electron states. These tail states remain globally localized and do not form an extended band. The extracted DOS therefore serves as a quantitative descriptor of energetic disorder and its composition dependence

We also extend the two-exponential tail fitting to mid-gap as shown in Fig. S10. The extracted density of states around 1.5eV near $10^{13}$ cm$^{-3}$, which is consistent with other study. The resulting DOS values are on the order of $10^{12}$–$10^{22}$ cm$^{-3}$ eV$^{-1}$ for the energy range 0-2 eV, which are consistent with previously reported values in similar disordered semiconductors.

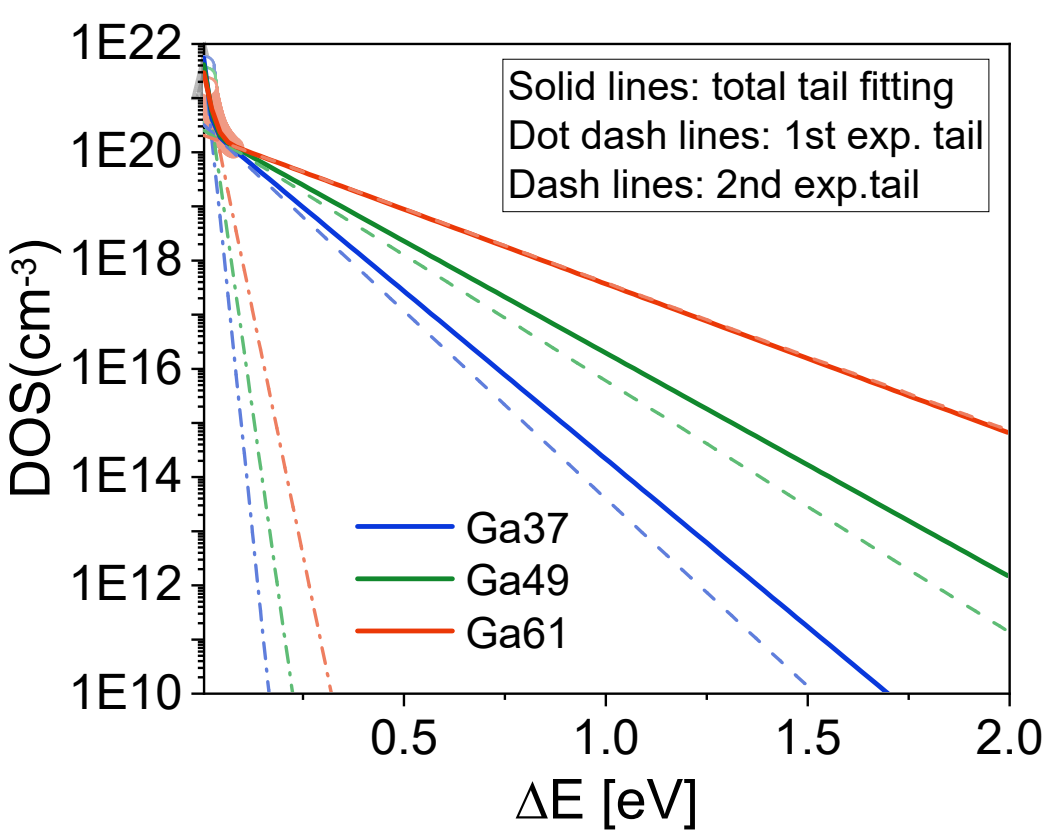

**Fig. S10** The extracted tail DOS of three compositions. The solid-color lines represent two-exponential tail fittings. Dash-dot lines correspond to the first exponential tail ($E_{0,1}$, $N_{0,1}$), while dashed lines represent the second tail ($E_{0,2}$, $N_{0,2}$).

**Table S2 The extracted tail widths and density**

| | $E_{0,1}$ [eV] | $N_{0,1}$ [cm$^{-3}$eV$^{-1}$] | $E_{0,2}$ [eV] | $N_{0,2}$ [cm$^{-3}$eV$^{-1}$] |
|---|---|---|---|---|
| Ga37 | 0.00525±5e-4 | 1.488e22 | 0.06165±0.0102 | 4.661e20 |
| Ga49 | 0.00726±9.05e-4 | 8.93e21 | 0.9343±0.0092 | 3.853e20 |
| Ga61 | 0.01215±0.001 | 3.552e21 | 0.15883±0.012 | 1.178e20 |

# Part 3 TCAD simulation for three a-IGZO compositions

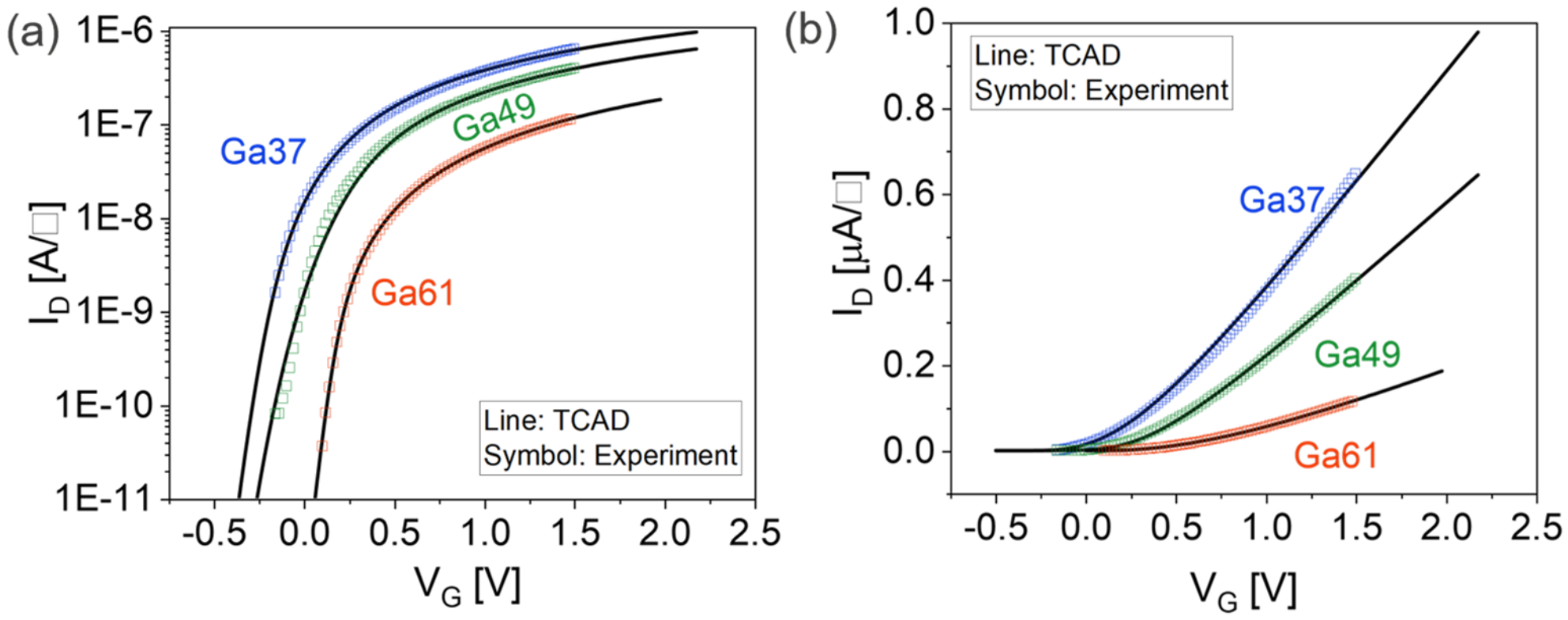

**Fig. S11** TCAD simulations using the extracted tail parameters from the density of states (DOS) in Fig. 2b. Simulation results are plotted on a logarithmic scale in (a) and a linear scale in (b)

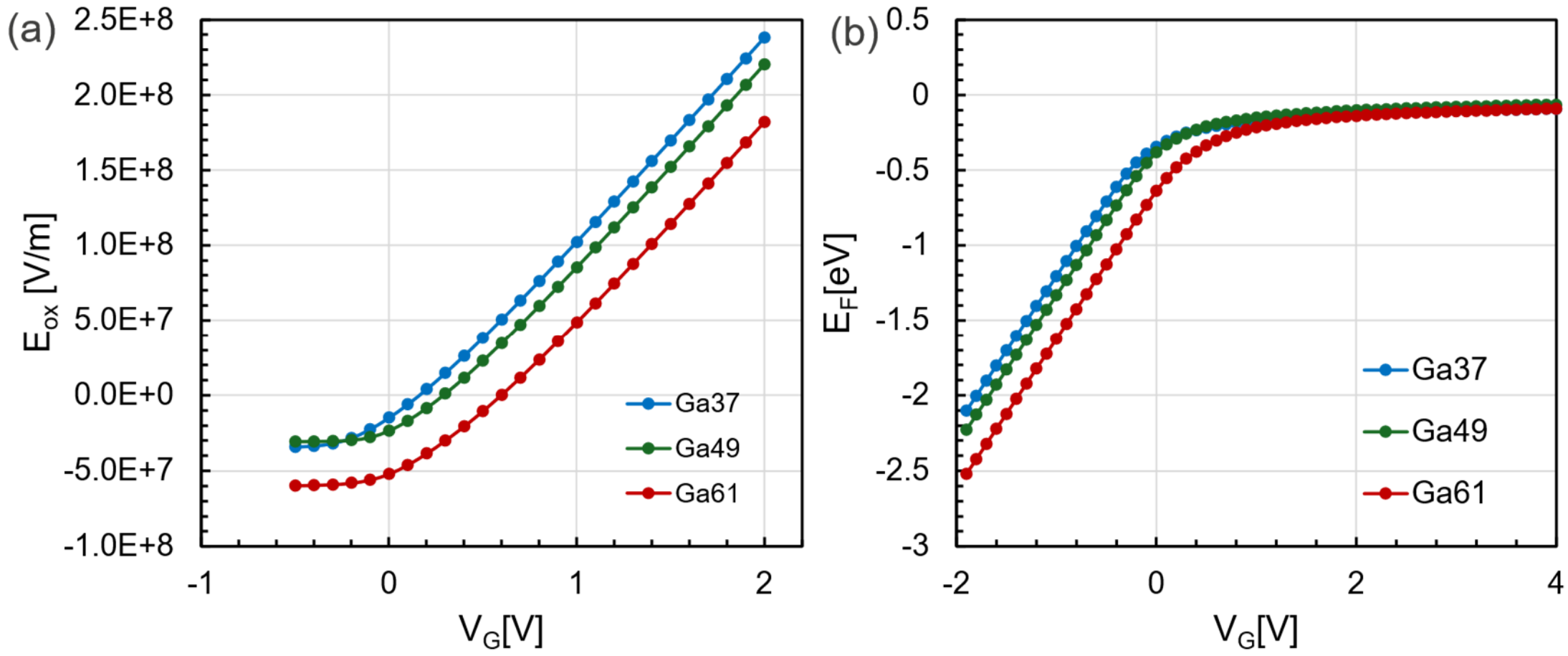

**Fig. S12** (a) The extracted oxide electric field ($E_{ox}$) from TCAD simulation for three a-IGZO compositions as a function of Gate voltage. (b) The extracted fermi level ($E_F$) from TCAD simulation changing with applied gate voltage. The reference 0 potential is set as the fully delocalized states ($E_{dloc}$). The separation $\phi$ between $E_F$ and $E_{dloc}$ is calculated by $E_{dloc}$ - $E_F$. Our simulations show that $E_F$ does not rise above $E_c$, even under large gate voltages. This result supports the picture of our transport model, where the electron filling level remains below $E_{dloc}$ and no global percolation path is formed.

## Note S2 $V_{FB}$ extraction

Simulations of the transfer characteristics are performed using Synopsys® Sentaurus to evaluate the flat-band voltage ($V_{FB}$) of the fabricated a-IGZO transistors. An equivalent two-dimensional device structure including a $p^{++}$ Si gate, 1nm $SiO_2$, 5 nm $Al_2O_3$, and 12 nm a-IGZO stacks is simulated. Material parameters including dielectric constants, work functions and electron affinities (based on internal photoemission (IPE) [3]) are defined in Table S1. For IGZO, conduction band tail states are implemented as exponentially distributed (in energy) traps, uniformly distributed (in space) in the whole channel. Higher IGZO bandgap value with higher Ga/Zn ratio is accounted for.

A quasi-static gate voltage sweep is simulated, and linear operating condition is ensured ($V_D = 0.05$ V). The electric field $E_{ox}$ across the gate dielectric is extracted at the $Al_2O_3$/a-IGZO interface.

By plotting $E_{ox}$ as a function of $V_G$, the $V_{FB}$ is identified as the gate voltage at which $E_{ox}$ becomes zero, corresponding to the condition where no net charge is present in the semiconductor.

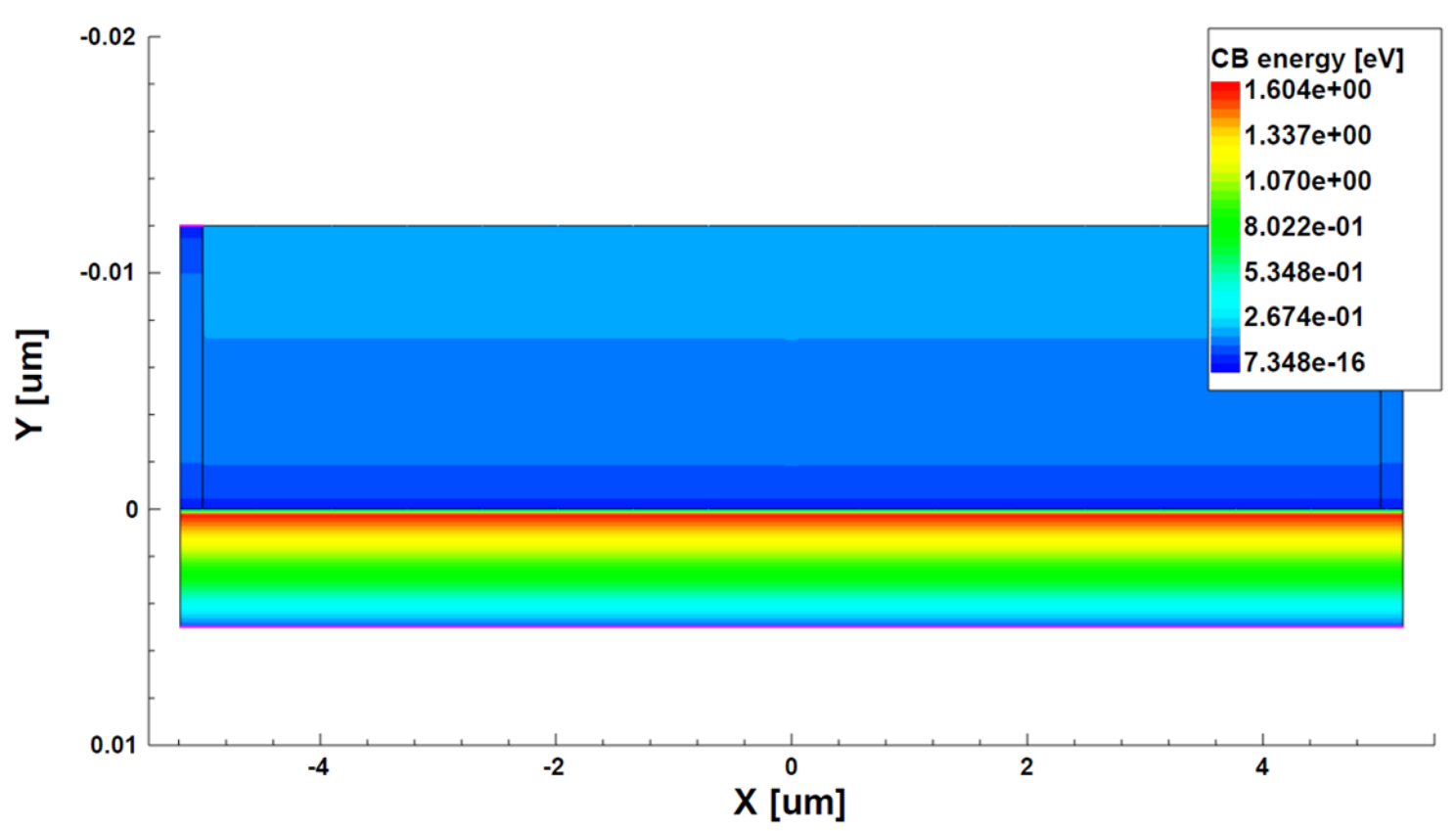

**Fig. S13** The equivalent two-dimensional device structure with the distribution of conduction band energy.

**Table S2. TCAD Parameters**

| Parameters | | Materials | | | | | |
|---|---|---|---|---|---|---|---|
| Name | Unit | Si | $SiO_2$ | $Al_2O_3$ | a-IGZO [In:Ga:Zn] | | |
| | | | | | 39:37:24 | 39:49:12 | 39:61:0 |
| $E_G$ | [eV] | 1.1 | 9.0 | 6.2 | 3.3 | 3.4 | 3.9 |
| $\chi$ | [eV] | 4.05 | 0.9 | 1.9 | 4.7 | 4.6 | 4.1 |
| $\epsilon_r$ | | 11.7 | 3.9 | 8 | 15 | 15 | 15 |

| Parameters | | a-IGZO stoichiometry [In:Ga:Zn] | | |
|---|---|---|---|---|
| Name | Unit | 39:37:24 | 39:49:12 | 39:61:0 |
| $n_{dop}$ | [$cm^{-3}$] | 1.26e+18 | 1.12e+18 | 2.20e18 |
| $N_{0_1}$ | [$cm^{-3}eV^{-1}$] | 1.98e+21 | 1.10e+21 | 1.00e21 |
| $E_{0_1}$ | [eV] | 0.004 | 0.005 | 0.01 |
| $N_{0_2}$ | [$cm^{-3}eV^{-1}$] | 3.99e+20 | 2.66e+20 | 9.29e19 |
| $E_{0_2}$ | [eV] | 0.070 | 0.090 | 0.120 |

# Part 4 Raw data for Hall measurements and extraction of Hall parameters

## Note S3: The fabrication of a-IGZO FETs with Hall-bar structure

The samples of a-IGZO FETs with Hall-bar structure were prepared from stacks deposited in 300 mm FAB. The wafers comprised a Si substrate, a 15 nm $Al_2O_3$ gate oxide, 10 nm a-IGZO channel layer, and a 100 nm $SiO_2$ capping layer. The wafers were cleaved into ~2.5 cm × 2.5 cm coupons for processing.

Device processing began with photolithography to define the active area. A 1 μm layer of positive photoresist (IX845) was spin-coated at 4000 rpm, soft-baked at 120 °C, and exposed using a Karl Suss MA6 aligner (400 nm UV, 3.5 s, hard contact, ligt-field mask). The samples were developed in OPD (o-phenylenediamine dihydrochloride) for 90 seconds and then rinsed in DIW (deionized water).

After the photoresist development, the top $SiO_x$ was dry etched (Oxford PlasmaPro 80 RIE, $CF_4$/Ar 30/10 sccm, 200 mTorr, 200 W, 180 s). Photoresist was stripped using heated Techni-strip in an ultrasonic bath (70°C, 80kHz, 10 min), followed by rinsing in IPA for 3 min and drying with nitrogen gun. Then, the a-IGZO layer was wet etched with oxalic acid (with top SiO2 acting as a mask), with an etch rate of 0.2 nm/s.

A second lithography (identical process to the first lithography) defined the contact trench, followed by a repeat of the above-described dry etch and photoresist removal steps. Next, the samples were annealed in dry air at 250 °C for 1 hour.

A third lithography step defined metal contact regions. A brief dry etch cleaned the top a-IGZO interface from potential oxide residuals. This was followed by Ti (10 nm) and Au (200 nm) deposition via sputtering and thermal evaporation. Metal lift-off was performed using ultrasonic sonication and pipetting to remove excess metal. A final annealing at 250°C for 1 hour completed the process.

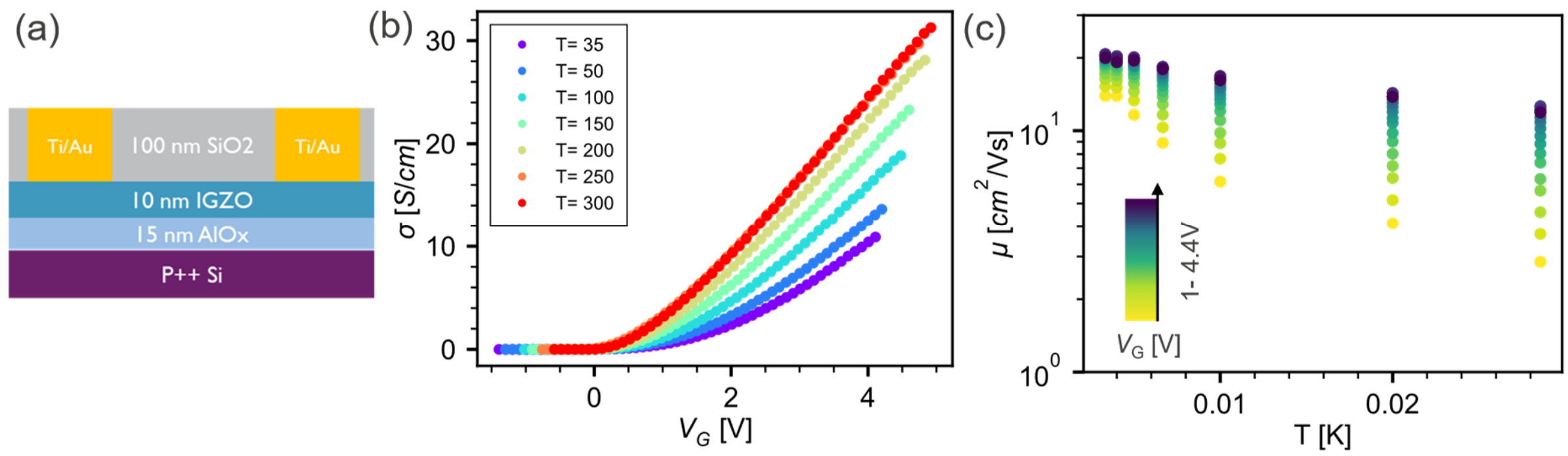

**Fig. S14** (a) Device structure of a-IGZO FETs with Hall-bar structure. (b) Transfer characteristics of Hall-bar devices under different temperatures, plotted in linear scale. (c) The extracted filed effect mobility of Hall devices as a function of the inverse of temperature (1/*T*).

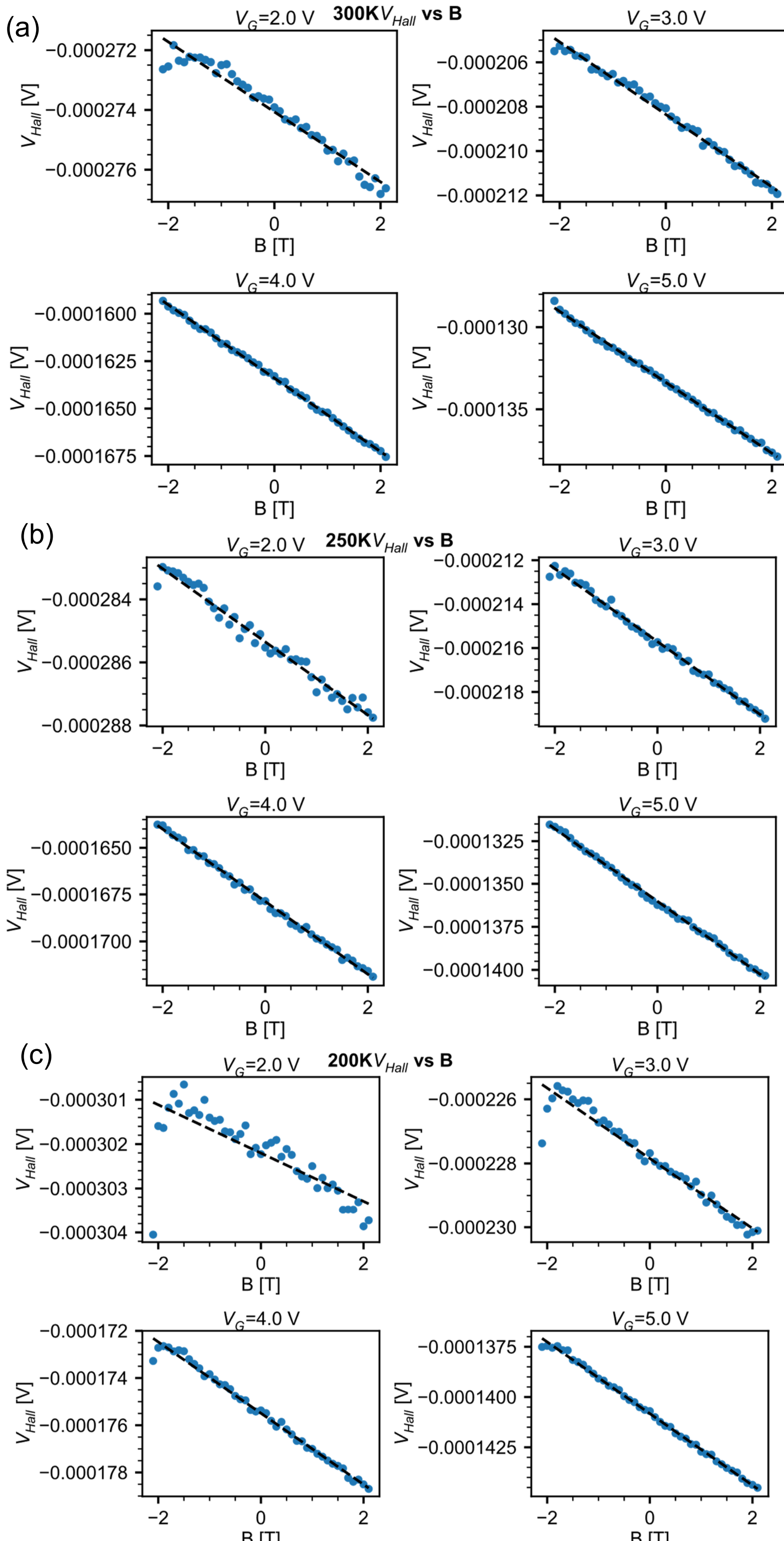

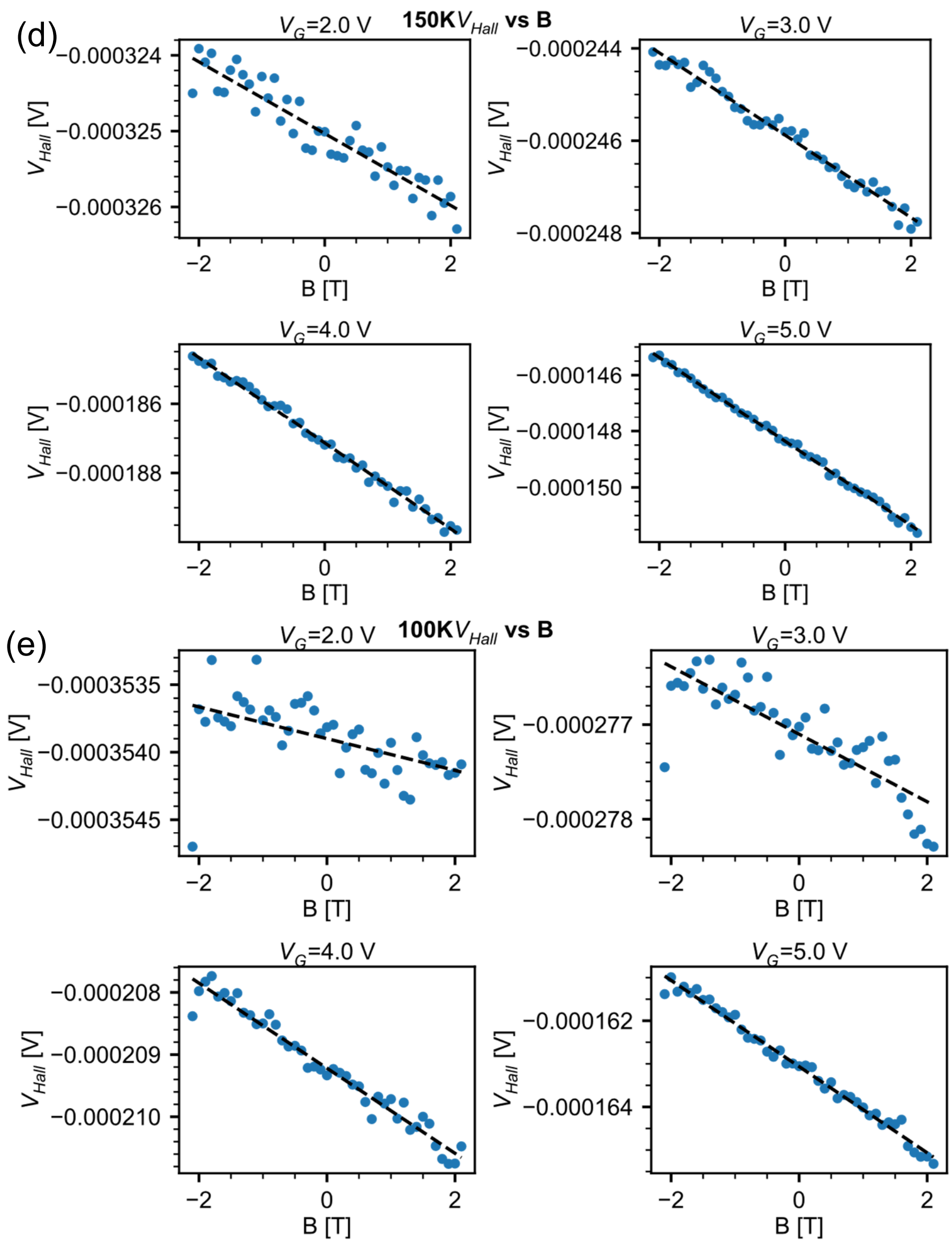

**Fig. S15** Measured Hall voltage $V_{\mathrm{Hall}}$ as a function of magnetic field $B$ under 300 K (a), 250 K (b), 200 K (c), 150 K (d) and 100 K (e). The dashed black lines are the linear fitting.

## Note S4: The extraction of Hall coefficient and Hall carrier concentration.

The Hall coefficient, $R_{\text{Hall}}$, was obtained from the slope $V_{\text{Hall}}(B)$ at fixed $V_{\text{G}}$ as

$$1/R_{\text{Hall}} = \frac{I_{\text{XX}}\,B}{V_{\text{Hall}}}, \tag{S2}$$

where $I_{\text{XX}}$ is the longitudinal current, $B$ is the magnetic field. The Hall carrier density $n_{\text{Hall}}$ was calculated as

$$n_{\text{Hall}} = \frac{1}{q\,R_{\text{Hall}}}, \tag{S3}$$

with $q$ the elementary charge. The Hall mobility was determined from

$$\mu_{\text{Hall}} = \frac{\sigma}{q\,n_{\text{Hall}}}, \tag{S4}$$

where $\sigma = \frac{I_{XX}\,L}{V_{XX}\,W}$, $L$ is the longitudinal probe spacing, $W$ the channel width and $V_{\text{XX}}$ is the longitudinal voltage drop. The longitudinal conductivity $\sigma$ was determined from 4-probe measurements as shown in Methods.

In parallel, the total field-effect carrier density $n_{\text{FET}}$ was obtained from Gaussian law

$$n_{\text{FET}} = C_i\,(V_{\text{G}} - V_{\text{th}})/q, \tag{S5}$$

where $C_i$ is the gate-channel capacitance (with equivalent dielectric permittivity $\varepsilon_i = \varepsilon_{SiO_2} = 3.9$, and $t_i$ = CET measured experimentally in Fig. S1.), where $V_{\text{th}}$ was defined from the extension of the slope of $1/R_{\text{Hall}}$ ($V_{\text{G}}$) to specify the threshold for inducing well-mobile carriers while disregarding trapped charge. The field-effect mobility in the linear regime was

$$\mu_{\text{FET}} = \frac{\partial I_{\text{D}}/\partial V_{\text{G}}\cdot L}{W\,V_{\text{D}}\,C_{\text{i}}}. \tag{S6}$$

In this work, $R_{\text{Hall}}$ and $n_{\text{Hall}}$ were evaluated in 2D sheet units, as Gauss's law in this context yields a sheet charge density. This choice enables a direct and accurate comparison between $n_{\text{Hall}}$ and $n_{\text{FET}}$. For comparison with literature data reporting bulk $n_{\text{Hall}}$, the sheet density values were converted to volume density (see Fig. S16). Comparison of $n_{\text{Hall}}$, $n_{\text{FET}}$, $\mu_{\text{Hall}}$, and $\mu_{\text{FET}}$ was interpreted in the context of the mixed band–hopping transport framework proposed Yi *et al.* [4]. Within this model, incoherent carriers involved in hopping conduction can partially compensate the Hall voltage, particularly in the strong-hopping regime, resulting in $n_{\text{Hall}} > n_{\text{FET}}$ and $\mu_{\text{Hall}} < \mu_{\text{FET}}$.

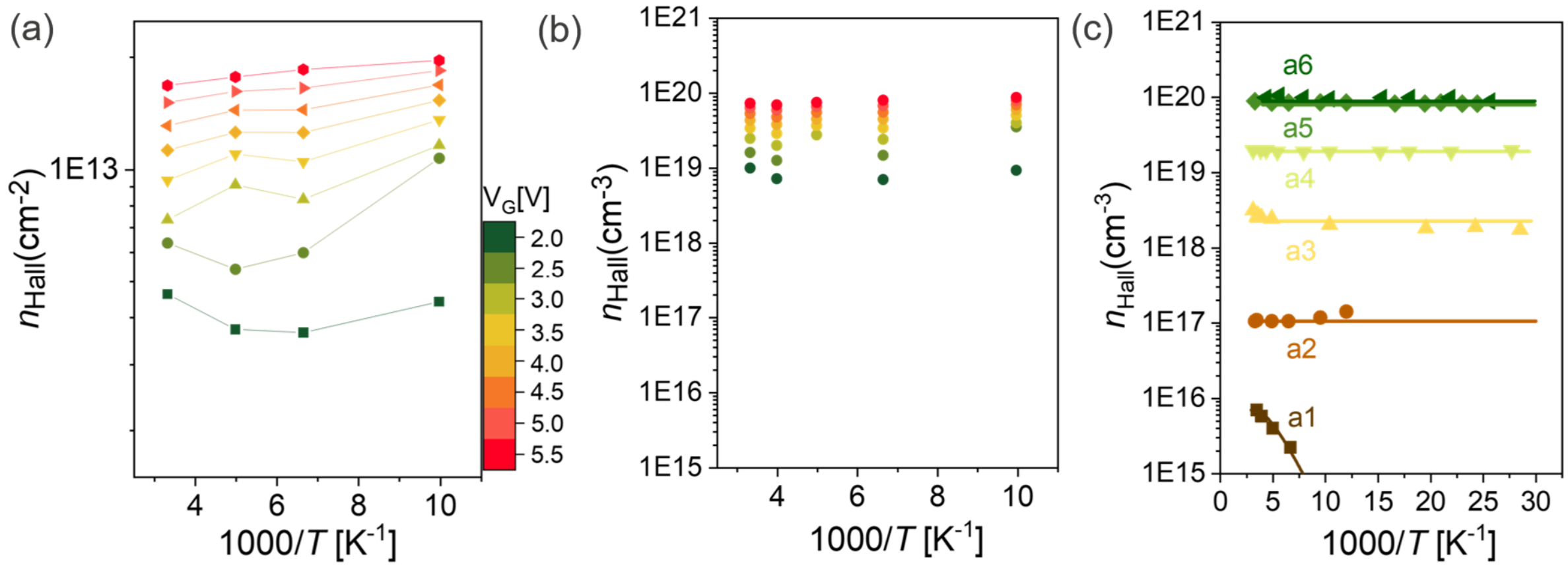

**Fig. S16** Hall carrier concentration per area (a) and per volume (b), extracted from our field effect Hall measurement as a function of the inverse of temperature (1/$T$); (c) Hall carrier concentration of doping a-IGZO as a function of 1/$T$ from paper [5]. a1-a5 means different doping level of a-IGZO.

The temperature independence of $n_{\text{Hall}}$ for a-IGZO with higher doping level is the powerful experimental proof in [5] of electrons transport degenerately. We notice that, if we plot the temperature dependence of our $n_{\text{Hall}}$ in the same range of y-axis as $n_{\text{Hall}}$ in [5], we see our $n_{\text{Hall}}$ also shows "temperature independence". However, if we plot $n_{\mathrm{Hall}}$ in a proper y-axis scale, the temperature dependence of $n_{\mathrm{Hall}}$ is obvious, see Fig. S16 (a). Also, from our results of field effect Hall measurements in the main paper, the comparison between $1/R_{\mathrm{Hall}}$ and $Q_{\mathrm{FET}}$ shows the contribution and compensation from incoherent electrons in a-IGZO, see Fig. 2d and Fig. 2e in the main text. Thus the "independence" of $n_{\mathrm{Hall}}$ on temperature is not the proof of that $E_{\mathrm{F}}$ goes above $E_{\mathrm{dloc}}$ and electrons have degenerate transport.

## Part 5 The atoms model of three compositions in first principles computations

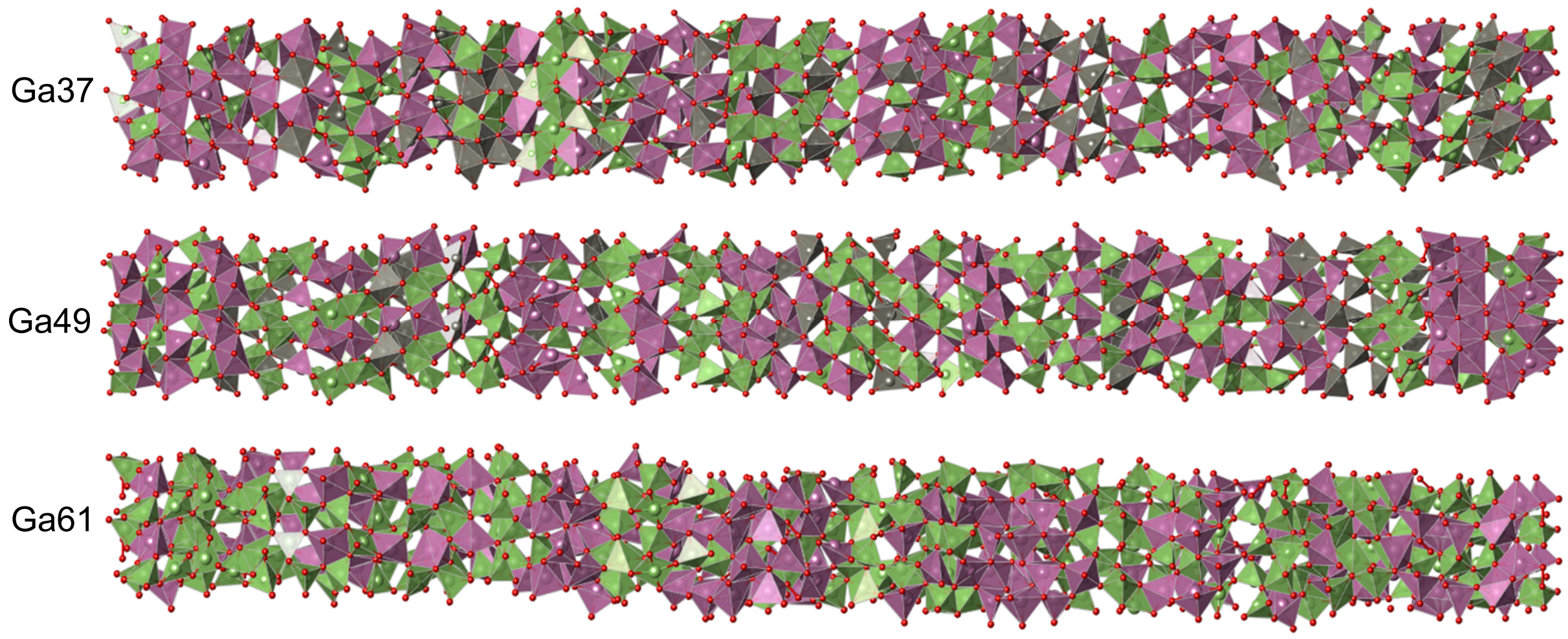

**Fig. S17** The elongated models of about 500 atoms in first principles computations.

# Part 6 Mathematical details of development of the FEAFIT model for a-IGZO TFTs

## Note S5 Development of the FEAFIT model for a-IGZO TFTs

In the main text we argue that transport in a-IGZO is governed by tunneling between locally coherent conducting puddles separated by thin insulating gaps, in the presence of thermal voltage fluctuations across those gaps. This is the physical content of the "dry lake-bed" picture: electrons are quasi-free inside puddles, but long-range conduction is limited by tunneling through the intervening barriers.

Mathematically, our formulation follows the fluctuation-induced tunneling (FIT) framework introduced by Sheng for disordered systems [6], but is explicitly adapted here to amorphous oxide semiconductors. We (i) apply the FIT mechanism to the a-IGZO dry-lake-bed energy landscape, where partially coherent conduction pockets naturally form nanoscale tunneling junctions, (ii) embed the junction in a field-effect geometry so that the effective barrier height and width acquire a controlled gate-bias dependence, and (iii) relate the resulting FIT parameters to composition-dependent a-IGZO quantities such as DOS-tail widths and DFT-derived pocket separations.

The goal of this Note is to show how, starting from a standard Landauer–type tunnelling picture and a WKB transmission through the image-force-corrected barrier, thermal voltage fluctuations across the small junction naturally lead to the compact conductivity expression used in the main text, and how the characteristic scales $T_1$ and $T_1/T_0$ relate to the microscopic barrier height $U$ and width $w$.

### S5.1 Barrier profile and image-force correction

We model the barrier separating two conducting puddles as a one-dimensional potential $U(x)$ of width $w$ and height $U$. A purely rectangular barrier would give unphysical infinite fields at its edges. Following Sheng [6], we apply an image-force correction to obtain a smooth barrier profile:

$$U(\bar{x}, \bar{\mathbb{E}}) = U_0 \left[1 - \frac{\lambda}{\bar{x}(1-\bar{x})} - \bar{\mathbb{E}}\bar{x}\right], \tag{S7}$$

where $\bar{x} = \frac{x}{w}$ is the reduced spatial variable, $0 < \bar{x} < 1$, $\bar{\mathbb{E}} = \mathbb{E}\frac{ew}{U_0}$ is the normalized electric field cross the junction, $U_0$ is the initial tunneling barrier height without correction, $e$ is electronic charge and $\lambda$ is a dimensionless parameter controlling the strength of the image-force rounding and relating to permittivity constant of semiconductor [6]. Larger values of $\lambda$ corrects the barrier shape more. $\lambda = 0.2$ makes the barrier more rounded and corrects the barrier from $U_0 = 0.5$ V to less than 0.1 V, as shown in Fig. S18(c). We can generate a variety of temperature variations interpolating between the thermally activated behavior at high temperatures and temperature-independent tunneling at low temperatures, by changing the value of $\lambda$ to equivalent varying the barrier shape. Barriers nearly rectangular in shape with small $\lambda$ yield curves more slowly approaching the low-temperature plateau, while more rounded barriers give in the opposite direction, as shown in Fig. S18(d).

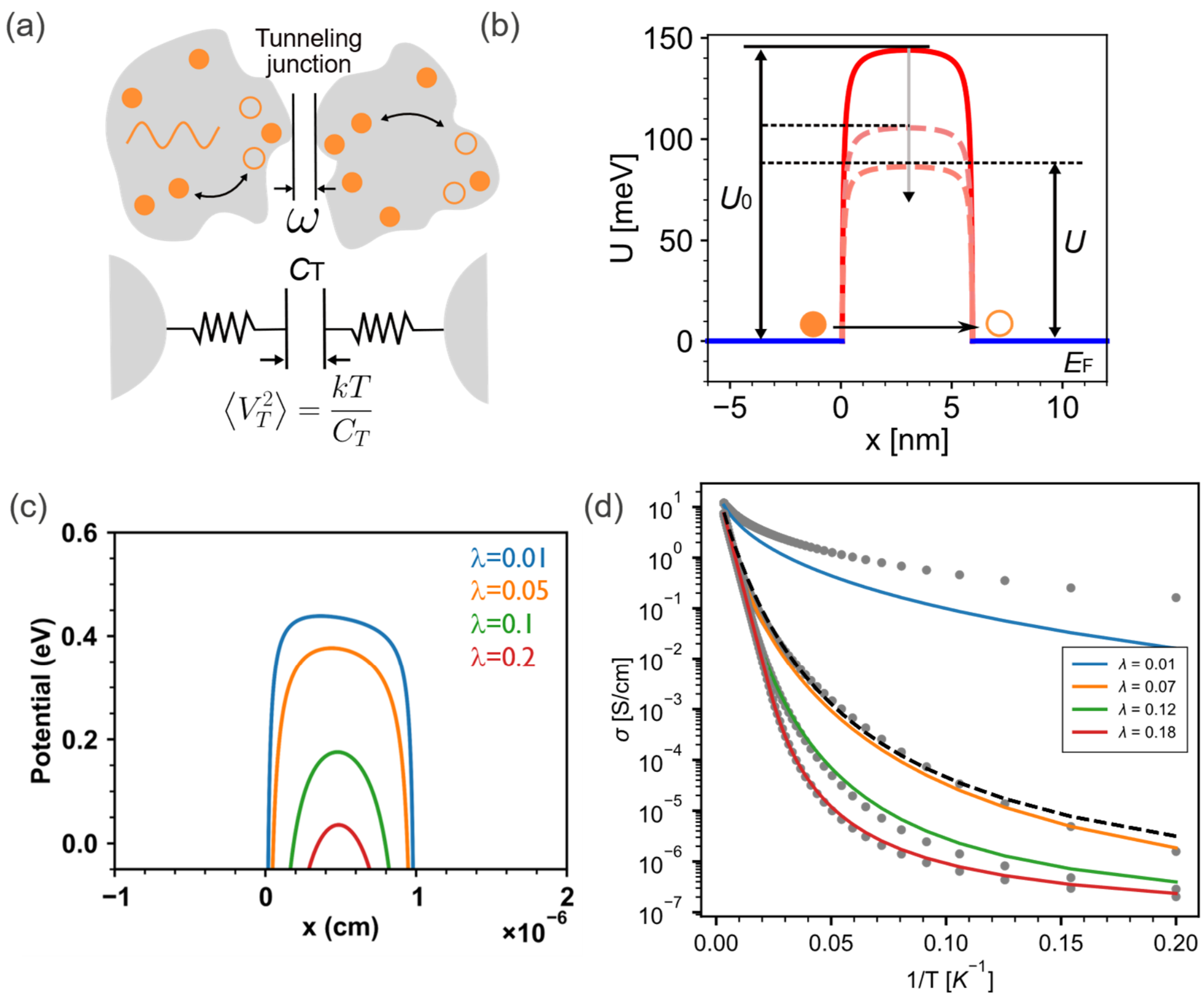

**Fig. S18** (a) Schematic of tunnel junction between two conducting regions in FIT model. The tunnel junction with width $w$ is modeled as a parallel-plate capacitor $C_T$. (b) Simulated reduction of the tunneling barrier due to temperature-induced voltage fluctuations in our model. Simulation results of potential barrier (c) and temperature dependence of conductivity (d) across tunneling junction with varying barrier shapes (controlled by $\lambda$), based on FIT model. The black dash line is the conductivity calculated by parabolic barrier shape. The gray symbols are the conductivity calculated by the numerical $\psi(\eta)$.

For fixed $U_0$ and $\lambda$, a critical field $\mathbb{E}_0$ is defined as the field at which the barrier maximum vanishes, $U_m(\mathbb{E}_0) = 0$. Mathematically, $\mathbb{E}_0$ is obtained by solving

$$U(\bar{x}_m, \bar{\mathbb{E}}_0) = 0, \qquad \frac{dU(\bar{x}, \bar{\mathbb{E}}_0)}{d\bar{x}}\Big|_{\bar{x}_m} = 0, \tag{S8}$$

for the location $\bar{x}_m$ of the maximum and the corresponding reduced critical field $\bar{\mathbb{E}}_0$. Substituting $\bar{\mathbb{E}}_0$ back into Eq. (S7) verifies that $U(\bar{x}_m, \bar{\mathbb{E}}_0) = 0$. Fig. S19(a) shows the dependence of $\mathbb{E}_0$ on $\lambda$: for $\lambda \gtrsim 0.25$, $\mathbb{E}_0$ becomes negative and the barrier maximum is already below zero even at zero applied field, which we regard as unphysical in our context. We therefore restrict $\lambda \leq 0.25$. Fig. S19 (b-f) show that under a calculated $\mathbb{E}_0$, $U_m$ is zero for different $\lambda$ and $U_0$.

It is convenient to express the fluctuating field through the dimensionless reduced field

$$\eta = \bar{\mathbb{E}}/\bar{\mathbb{E}}_0, \qquad 0 \leq \eta \leq 1, \tag{S9}$$

In this parametrization, $\eta = 0$, corresponds to zero field and $\eta = 1$ corresponds to the critical field at which the barrier maximum is driven to zero. Fig. S19(b)-(f) illustrate the barrier profile $U(\bar{x}, \eta)$ for different $\lambda$ and $U_0$. At $\eta = 0$ ($\bar{\mathbb{E}}_0$=0) the image-force-corrected barrier is finite and smooth, while increasing $\eta$ lowers and narrows the barrier, and at $\eta = 1$ ($\bar{\mathbb{E}} = \bar{\mathbb{E}}_0$) the top of the barrier reaches zero.

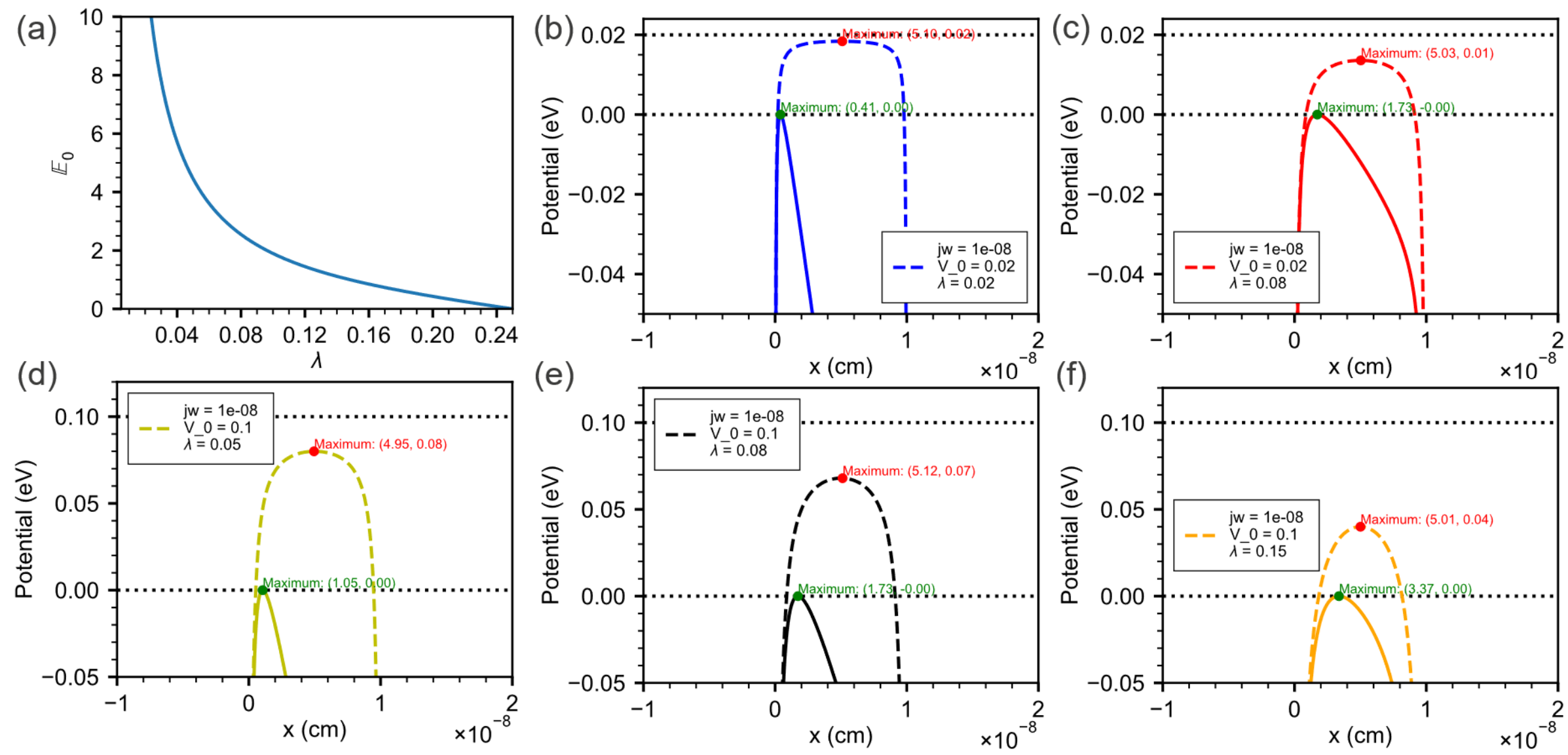

**Fig. S19** (a) Variation of $\mathbb{E}_0$ as a function of $\lambda$. (b)-(e) The tunneling barriers with different $\lambda$ and barier height $U_0$. The black dot lines are the energy zero and the $U_0$. The dashed lines are the barrier shapes with image force correction. The solid barrier shape curves show the barrier under critical filed $\mathbb{E}_0$, $\eta = 1$.

## S5.2 Landauer-type tunnelling conductance and WKB

For a fixed internal field $\eta$ across the tunneling barrier, the conductance of a single junction can be written in Landauer form[7]

$$G(\eta) = \frac{2e^2}{h} \int dE\, \mathcal{T}(E,\eta)\, M(E)(-\frac{\partial f}{\partial E}), \tag{S10}$$

where $\mathcal{T}(E,\eta)$ is the transmission probability through the barrier, $M(E)$ is the number of transverse modes, and $f(E)$ is the Fermi function. In the small-bias, low-temperature limit, the derivative $-\partial f/\partial E$ is sharply peaked around the Fermi level $E_F$, so that

$$G(\eta) \propto \mathcal{T}(E_F,\eta), \tag{S11}$$

Thus, for a static barrier, the conductance is essentially proportional to the tunnelling transmission evaluated at the Fermi level. We absorb the prefactor into $\sigma_0$ in what follows.

Within the WKB approximation [8], the transmission probability through a one-dimensional barrier is written in terms of the WKB action $S(E,\eta)$ in the classically forbidden region as

$$\mathcal{T}(E,\eta) \approx \mathcal{T}_0(E,\eta) \exp\left[-\frac{2}{\hbar} S(E,\eta)\right], \tag{S12}$$

with

$$S(E,\eta) = \int_{x_1}^{x_2} \sqrt{2m_e(U(x,\eta) - E)}\, dx, \tag{S13}$$

where $m_e$ is the effective electron mass and $x_1, x_2$ are the classical turning points defined by $U(x,\eta) = E$. The prefactor $\mathcal{T}_0(E,\eta)$ contains the usual WKB normalization (often involving factors such as $1/S(E,\eta)$ in textbook expressions), but it varies only weakly with temperature compared to the exponential term and is therefore absorbed into the overall prefactor $\sigma_0$ in our conductivity expression. For notational convenience we collect the factor $2/\hbar$ into a dimensionless exponent

$$F(E,\eta) = \frac{2}{\hbar} S(E,\eta), \tag{S14}$$

and write

$$\mathcal{T}(E,\eta) \approx \exp\left[-F(E,\eta)\right]. \tag{S15}$$

For a-IGZO transistor, we take the Fermi level of the whole transistor as our energy reference, $E_F = 0$, and approximate the action at $E \simeq 0$. Inserting the image-force-corrected barrier (Eq. S7) and expressing the integral in reduced coordinates gives

$$F(0,\eta) = 2\chi w\,\xi(\eta), \chi = \sqrt{\frac{2m_e U_0}{\hbar^2}}, \tag{S16}$$

with the dimensionless action

$$\xi(\eta) = \int_{\bar{x}_1}^{\bar{x}_2} \sqrt{1 - \frac{\lambda}{\bar{x}(1-\bar{x})} - \eta\bar{\mathbb{E}}_0\bar{x}}\, d\bar{x}, \tag{S17}$$

where $\bar{x}_1, \bar{x}_2$ are the reduced turning points and $\bar{\mathbb{E}}_0$ is the normalized critical field introduced above. In these terms the static (no-fluctuation) conductance obeys

$$G(\eta) \propto \exp\left[-2\chi w\,\xi(\eta)\right]. \tag{S18}$$

It is useful to factor out the zero-field geometry by defining the normalized tunnelling action

$$\psi(\eta) = \frac{\xi(\eta)}{\xi(0)}, \tag{S19}$$

which measures how much the barrier is effectively reduced by the internal field. By construction, $\psi(0) = 1$ and, at the critical field $\eta = 1$, $\psi(1) = 0$ because the barrier maximum vanishes and the forbidden region collapses.

For the image-force-corrected barrier, equations (S17) and (S19) cannot be evaluated in closed form. Sheng provided an accurate analytic approximation,

$$\psi(\eta) \simeq \frac{1-\eta}{1+\alpha\eta+\beta\eta^2}, \tag{S20}$$

where $\alpha$ and $\beta$ depend on $\lambda$. We have checked this expression against direct numerical evaluation of $\xi(\eta)$. For the $\lambda$ values relevant to a-IGZO (λ (≳0.08)), as shown in Fig. S20, the analytic and numerical $\psi(\eta)$ almost overlap. The deviations at small $\lambda$ are minor and do not change $\sigma(T)$ within our accuracy. When the numerical $\psi(\eta)$ is used in the full conductivity expression, the resulting $\sigma(T)$ curves are indistinguishable from those obtained with Eq. (S20) (Fig. S18 (d)). We therefore use the analytic form for efficiency and for its correct limits at $\eta = 0$ and $\eta = 1$.

For comparison, a simple parabolic barrier would give $\psi(\eta) = (1-\eta)^2$, but this approximation fails to describe a-IGZO across all compositions. The image-force-corrected form with composition-dependent $\lambda$ is required.

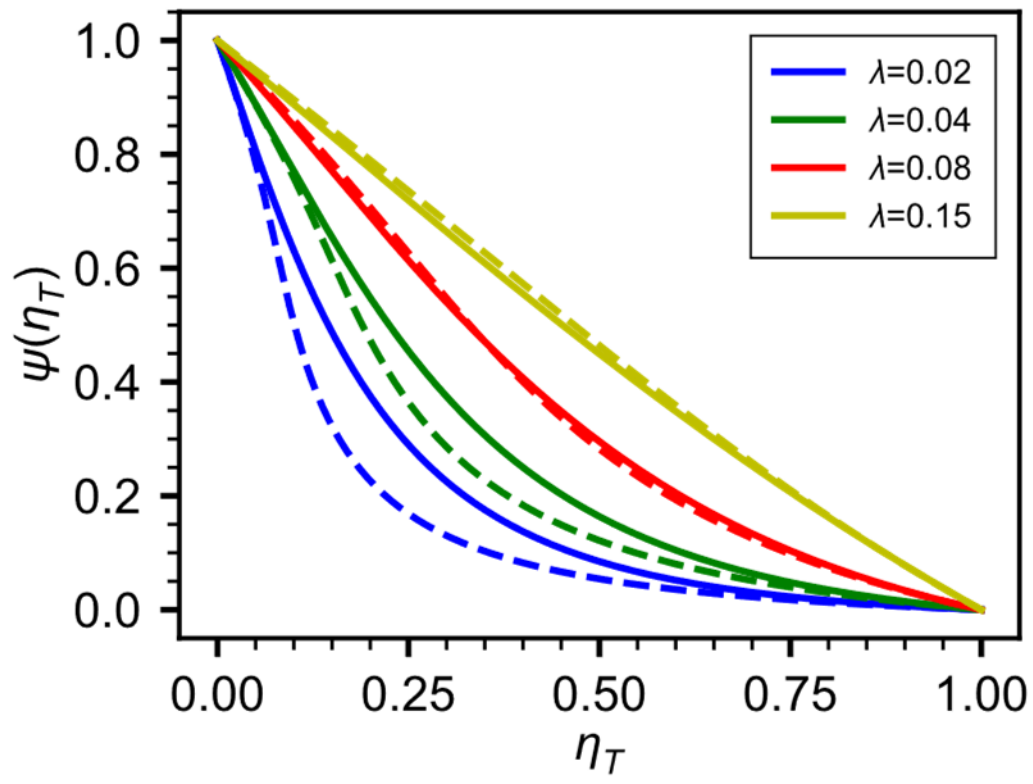

**Fig. S20** Variation of $\psi$ as a function of $\eta$. The dashed lines are from numerical solution and the solid lines are from the close form of Eq. (S20).

## S5.3 Thermal voltage fluctuations and distribution of the internal field

In nanoscale junctions, the electric field across the barrier is not strictly fixed but fluctuates due to the thermal charging and discharging of the gap between puddles. We model this gap as a parallel-plate capacitor of area $A$ and width $w$, with capacitance

$$C_T = \varepsilon \frac{A}{w}, \tag{S21}$$

where $\varepsilon$ is the permittivity of material. Equipartition implies that the mean-square voltage fluctuation across this capacitor is

$$\langle V_T^2 \rangle = \frac{kT}{C_T}, \tag{S22}$$

and, assuming Gaussian statistics, the probability density for a fluctuation $V_T$ is

$$P(V_T) \propto \exp\left[-\frac{C_T V_T^2}{2kT}\right]. \tag{S23}$$

The fluctuating electric field is $\mathbb{E}_T = V_T/w$. Using the reduced field $\eta = \mathbb{E}_T/\mathbb{E}_0$ (Eq. S9) and Eq. (S21), equation (S23) becomes

$$P(\eta) \propto \exp\left[-\frac{wA\varepsilon_0\mathbb{E}_0^2}{2kT}\,\eta^2\right] = \exp\left[-\frac{T_1}{T}\,\eta^2\right], \tag{S24}$$

with

$$T_1 = \frac{wA\varepsilon_0\mathbb{E}_0^2}{2k}. \tag{S25}$$

Thus the quadratic term $-T_1/T\,\eta^2$ in the FIT exponent directly reflects the Gaussian distribution of thermal voltage fluctuations across the nanoscale junction capacitor. $T_1$ sets the characteristic fluctuation energy scale.

At a given instantaneous internal field $\eta$, transport is described by the static Landauer–WKB conductance $G(\eta)$ in Eq. (S18). The observed conductivity is obtained by averaging this conductance over the fluctuating internal field:

$$\sigma(T) \propto \int_0^1 P(\eta)\, G(\eta)\, d\eta. \tag{S26}$$

Using Eqs. (S18) and (S24) and collecting prefactors, this becomes

$$\sigma(T) \propto \int_0^1 \exp\left[-\frac{T_1}{T}\,\eta^2 - \frac{T_1}{T_0}\,\psi(\eta)\right] d\eta, \tag{S27}$$

where

$$\frac{T_1}{T_0} = \frac{2\sqrt{2m_e}}{\hbar}\,\xi(0)\,w\sqrt{U} \tag{S28}$$

collects the WKB prefactors and barrier geometry. Equation (S25) and Eq. (S28) are equivalent to Sheng's definitions of $T_1$ and $T_0$, written here explicitly in terms of the junction width $w$, barrier height $U$, critical field $\mathbb{E}_0$, and zero-field action $\xi(0)$.

## S5.4 Saddle-point evaluation and most-probable fluctuation

The exponent in Eq. (S27),

$$\mathcal{F}(\eta, T) = -\frac{T_1}{T}(\eta)^2 - \frac{T_1}{T_0}\psi(\eta), \tag{S29}$$

contains two contributions. The fluctuation term $-\frac{T_1}{T}(\eta)^2$ penalizes large internal fields $\eta$, because such fluctuations are less probable. The tunnelling term $-\frac{T_1}{T_0}\psi(\eta)$ becomes less negative as $\eta$ increases, since a larger field lowers the barrier and reduces the WKB action. The combined function $\mathcal{F}(\eta, T)$ therefore increases with $\eta$ at small $\eta$ (barrier lowering wins), then decreases at larger $\eta$ (fluctuation penalty dominates), and exhibits a single maximum at an intermediate value $\eta^*(T)$. Figure S22(a) shows $-T_1/T\,\eta^2$, $-T_1/T_0\,\psi(\eta)$, and their sum $\mathcal{F}(\eta,T)$ as a function of $\eta$.

Because the integrand in Eq. (S27) is sharply peaked around this maximum, the integral can be evaluated by the saddle-point method. The most-probable normalized field $\eta^*(T)$is defined by

$$\frac{d\mathcal{F}}{d\eta}\Big|_{\eta^*} = 0 \Longrightarrow \frac{2\eta^*}{T} + \frac{1}{T_0}\,\psi'(\eta^*) = 0. \tag{S30}$$

We determine $\eta^*(T)$ numerically from this condition. For representative parameters, Figure S22(b–e) show how $\mathcal{F}(\eta,T)$ develops a pronounced maximum and how its position shifts with temperature $T_1$ and $T_0$ and $\lambda$. Figure S22(f) summarizes the resulting $\eta^*(T)$: at low temperature, $\eta^* \to 0$, as $T$ increases, $\eta^*(T)$ moves smoothly towards 1.

Evaluating Eq. (S27) at the saddle point then gives

$$\sigma(T) = \sigma_0 \exp\left[-\frac{T_1}{T}\,\eta^{*2} - \frac{T_1}{T_0}\,\psi(\eta^*)\right]. \tag{S31}$$

The two limiting regimes are immediately visible: for $T \to 0$, $\eta^* \to 0$ and $\psi(\eta^*) \to 1$, so that

$$\sigma(T \to 0) \to \sigma_0 \exp\left[-\frac{T_1}{T_0}\right], \tag{S32}$$

corresponding to nearly temperature-independent elastic tunnelling. For $T \to \infty$, the fluctuation penalty weakens, $\eta^* \to 1$ and $\psi(\eta^*) \to 0$, and one recovers a thermally activated Arrhenius form

$$\sigma(T \to \infty) \propto \exp\left[-\frac{T_1}{T}\right]. \tag{S33}$$

The continuous evolution of $\eta^*(T)$ from 0 at low $T$ to 1 at high $T$ thus explains the crossover from quasi-saturated tunnelling behaviour to thermally activated transport.

For compactness in the main text, it is convenient to define two temperature-dependent crossover factors,

$$S_1(T) = (\eta^*)^2,\ S_0(T) = \psi(\eta^*), \tag{S34}$$

and rewrite Eq. (S31) as

$$\sigma(T) = \sigma_0 \exp\left[-\frac{T_1}{T}\,S_1(T) - \frac{T_1}{T_0}\,S_0(T)\right]. \tag{S35}$$

In this form, $T_1$ and $T_1/T_0$ appear as fixed material-dependent scales, while $S_1(T)$ and $S_0(T)$ are smooth crossover functions that interpolate between the tunnelling and Arrhenius limits.

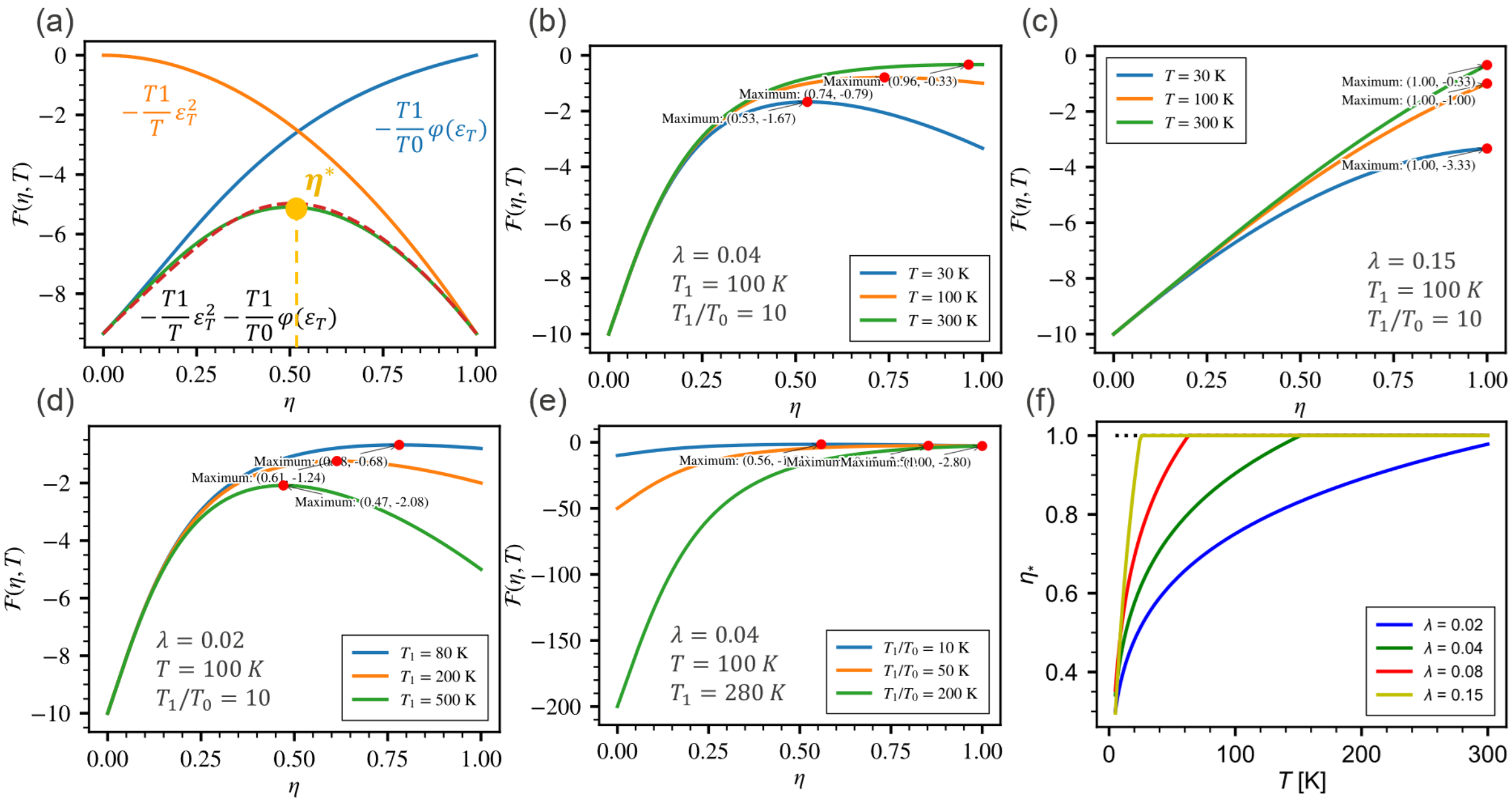

**Fig. S21** (a) The separate contribution $-\frac{T_1}{T}(\eta)^2$ and $-\frac{T_1}{T_0}\psi(\eta)$ and their sum $\mathcal{F}(\eta, T)$ as a function of normalized fluctuating electrical field $\eta_T$. (b)-(e) $\mathcal{F}(\eta, T)$ and the maximums with different $\lambda$, $T$, $T_1$ and $T_0$. (f) Variation of $\eta^*$ as a function of $T$.

## S5.5 Interpretation of $T_1$ and $T_1/T_0$ and gate-bias dependence

From Eq. (S25) and (S28), the characteristic scales $T_1$ and $T_1/T_0$ are directly linked to the barrier geometry and height. $T_1$ scales with $wA\mathbb{E}_0^2$ and sets the overall strength of the fluctuation term. It controls the effective activation energy in the high-temperature regime. The ratio $T_1/T_0$ scales with $\xi(0)\, w\sqrt{U_0}$ and determines how strongly the static barrier suppresses the low-temperature plateau of $\sigma$. Together, these two parameters provide complementary measures of the energetic disorder experienced by electrons moving between puddles.

In a TFT geometry, both the barrier height $U$ and width $w$ depend on the gate voltage. At zero gate bias, $U_0$ and $w_0$ are defined by the equilibrium Fermi level $E_{F0}$ and the composition-dependent potential landscape in a-IGZO. They describe the barrier between conduction puddles in the absence of band bending. Under a positive gate voltage $V_G$, the conduction band at the channel surface bends downward and the separation between $E_C$ and $E_F$ is reduced.

We obtain $U(V_G)$ from TCAD simulations of the TFT structure, which provide the gate-induced change in $E_C - E_F$ at the channel surface. We identify this energy difference with the effective barrier height between puddles at that $V_G$, i.e. $U(V_G) \simeq E_C(V_G) - E_F(V_G)$. For each $V_G$, the corresponding barrier width $w(V_G)$ is then determined from the image-force-corrected barrier profile (Eq. S7) by solving $U(\bar{x}, \eta) = E_F$ for the turning points $\bar{x}_1, \bar{x}_2$ and taking $w(V_G) \propto (\bar{x}_2 - \bar{x}_1)\, w_0$, where $w_0$ is the width at zero bias. In practice, the detailed prefactors are absorbed into the fitted parameters, and the important point is that both $U(V_G)$ and $w(V_G)$ decrease as $V_G$ increases.

Substituting $U(V_G)$ and $w(V_G)$ into Eqs. (S25) and (S28) gives $T_1(V_G)$ and $T_1(V_G)/T_0(V_G)$, which are then used in Eq. (S31). As the gate voltage increases, the barrier becomes lower and narrower, so $T_1$ and $T_1/T_0$ decrease with $V_G$. By fitting the measured $\sigma(T, V_G)$ over the full temperature range with Eq. (S31), we extract the initial barrier parameters $U_0$ and $w_0$ for each composition. These extracted values are compared in the main text with the DOS-derived tail widths and with barrier statistics from DFT, and show consistent trends with Ga content.

### S5.6 Model fitting strategy

For each composition and gate bias, the conductivity $\sigma(T, V_G)$ is evaluated from Eq. (S31-35) using the barrier parameters $U(V_G)$ and $w(V_G)$ obtained in Sections S5.3–S5.5. At $V_G = 0$, the quantities $w_0$ and $A$ characterize the intrinsic junction geometry of the a-IGZO landscape. Under applied gate bias, $U(V_G)$ is taken from TCAD-simulated band bending, while $w(V_G)$ is determined from the turning points of the image-force–corrected barrier (Sections S5.5). We perform a global fit of $\sigma(T, V_G)$ covering the full temperature range (10–300 K) and all measured gate voltages for each composition. The model parameters are extracted from the fitting. Across gate bias, the fitted quantities vary smoothly, reflecting the monotonic reduction of barrier height and width with increasing $V_G$. The resulting $U_0$ and $w_0$ show systematic composition dependence consistent with the DOS-tail widths and DFT-calculated separations between conduction pockets.

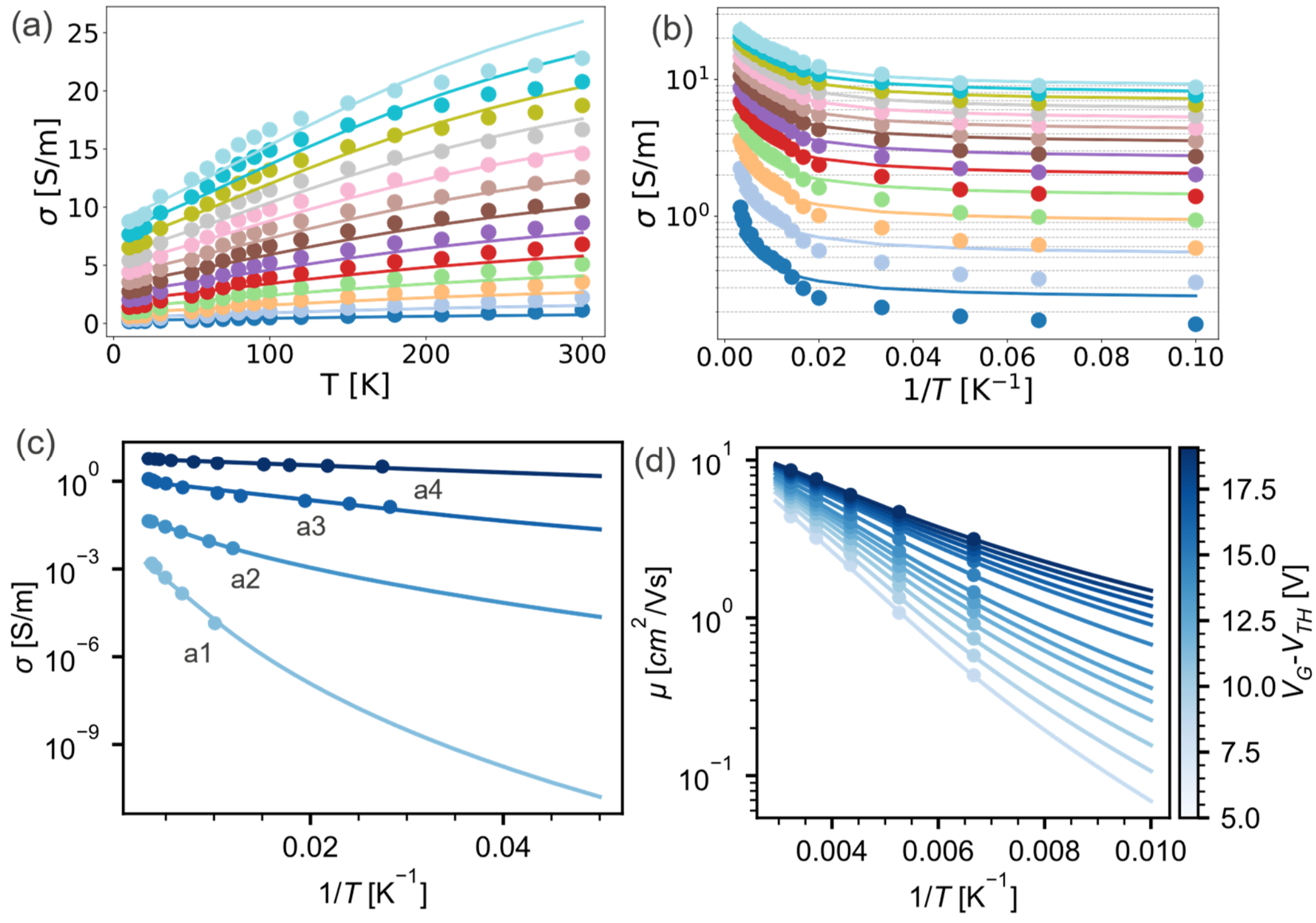

**Fig. S22** Comparison between our experimentally extracted conductivity ($V_D$=0.05 V) for Ga37 and the conductivity calculated by the model from paper of Kamiya[5], in linear scale (a) and logarithm scale (b). The model from paper of Kamiya[5] could generally match the trend of temperature dependence of conductivity but has obvious discrepancy under 100K. (c) The fitting results of experiments data in paper of Kamiya[5] by our FEAFIT. The labels a1-a4 is the different doping level defined in paper of Kamiya[5]. (d) The fitting results of experiments data in paper of Germs[9] by our FEAFIT. Since paper of Germs[9] only have several temperature points above 150K, we stop the fitting at 100K.

In amorphous IGZO, field-effect mobility is not a uniquely defined microscopic quantity. Because electronic states near the Fermi level are only locally coherent, the effective carrier density $n_{\mathrm{eff}}(T, V_G)$ depends sensitively on gate bias and temperature through the filling of the DOS tail. As $E_F$ moves to higher energies in the tail, the density of available states increases and the conducting regions expand, leading to an apparent increase in mobility. Consequently, mobility reflects a device-level average rather than an intrinsic transport parameter, whereas conductivity remains a well-defined observable that directly reflects the tunneling process described by the FEAFIT model. For completeness, the mobility is reconstructed from the calculated conductivity via $\mu(T, V_G) = \frac{\sigma(T,V_G)}{q\, n_{\mathrm{eff}}(T,V_G)}$, with $n_{\mathrm{eff}}(T, V_G)$ obtained from the corresponding field-effect conditions by fermi distribution and DOS.

To further examine the robustness and comparability of the FEAFIT parameters, we apply the same fitting procedure to conductivity data reported in the literature[5,9]. The FEAFIT can reproduce the data well, as shown in Fig.S22c-d. The extracted tunnelling parameters $T_1$ and $T_1/T_0$ can be obtained from temperature-dependent conductivity alone. When plotted as a function of conductivity at room temperature, both parameters of all the data fall on the same systematic trends. In particular, higher conductivity consistently correlates with lower $T_1$ and smaller $T_1/T_0$, indicating reduced effective barrier height and width. This observation demonstrates that the FEAFIT parameters are not arbitrary fitting variables, but can capture intrinsic aspects of the transport landscape in a-IGZO. It also confirms that the model provides a unified description of charge transport across different IGZO systems.

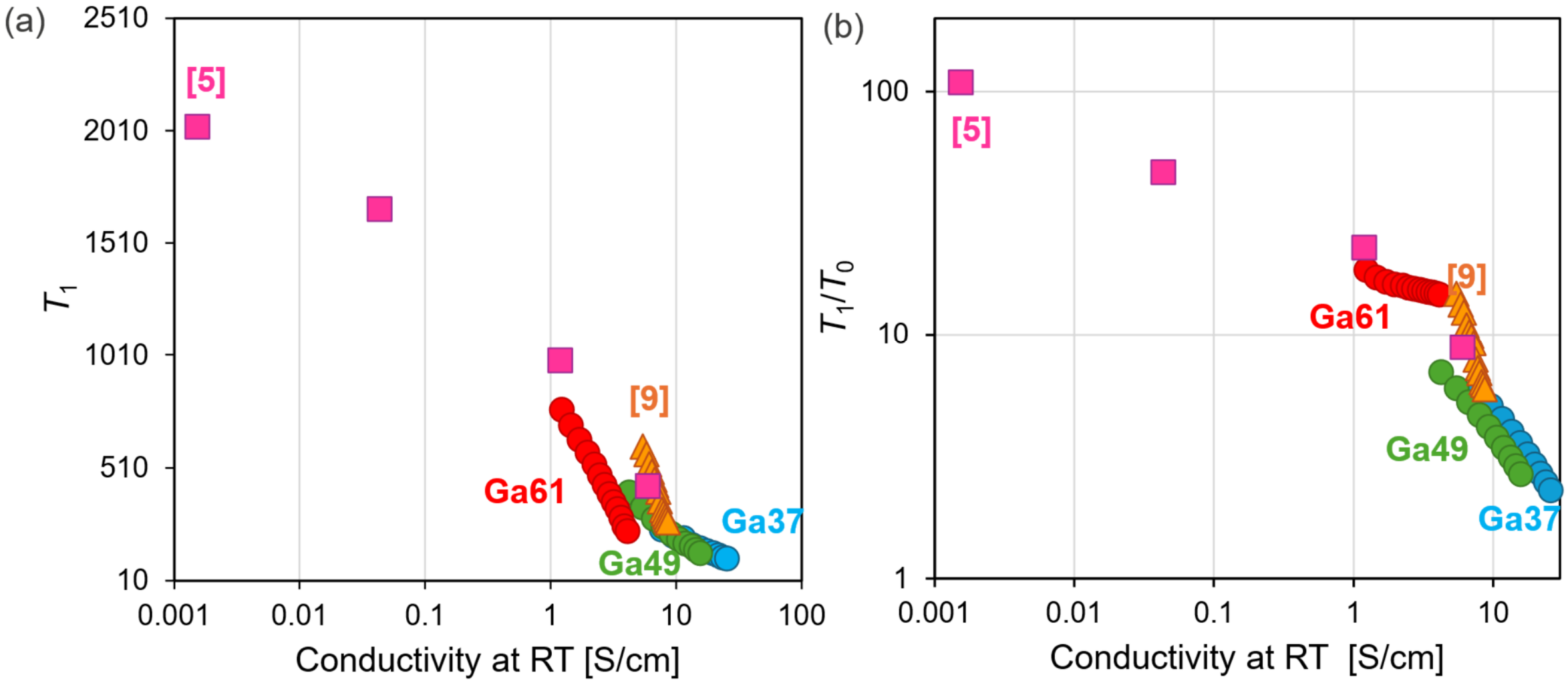

Fig. S23 FEAFIT-extracted tunnelling parameters $T_1$ (a) and $T_1/T_0$ (b) obtained from fitting results of experiments data in paper of Kamiya[5] and Germs[9], together with our a-IGZO devices with different Ga compositions (Ga37, Ga49, Ga61). The parameters are plotted as a function of conductivity at room temperature (RT) to illustrate their systematic evolution with macroscopic transport.

## Reference of Supplementary